\documentclass[twocolumn]{aastex631}

\usepackage{comment}
\usepackage{xcolor}
\usepackage{amsmath}

\newcommand{\pfrac}[2]{\left( \frac{#1}{#2} \right)}

\newcommand{\rp}{r_{\rm p}}

\newcommand{\rtidal}{r_{\rm t}}
\newcommand{\rh}{r_{\rm h}}

\newcommand{\tgw}{\tau_{\rm GW}}

\newcommand{\ttb}{\tau_{\rm 2B}}
\newcommand{\tdyn}{\tau_{\rm dyn}}
\newcommand{\MBH}{M_{\bullet}}

\newcommand{\kb}{k_{\rm B}}
\newcommand{\amt}{\alpha_{\rm MT}}
\newcommand{\ergss}{\rm erg \, s^{-1}}

\submitjournal{ApJ}

\begin{document}

\title{Tidal Disruption of a Star on a Nearly Circular Orbit}

\correspondingauthor{Itai Linial}
\email{itailin@ias.edu, il2432@columbia.edu}

\author[0000-0002-8304-1988]{Itai Linial}
\affil{Department of Physics and Columbia Astrophysics Laboratory, Columbia University, New York, NY 10027, USA}
\affil{Institute for Advanced Study, 1 Einstein Drive, Princeton, NJ 08540, USA}

\author[0000-0001-9185-5044]{Eliot Quataert}
\affiliation{Department of Astrophysical Sciences, Princeton University, Peyton Hall, Princeton, NJ 08540, USA}

\begin{abstract}
We consider Roche lobe overflow (RLO) from a low-mass star on a nearly circular orbit, onto a supermassive black hole (SMBH). If mass transfer is unstable, its rate accelerates in a runaway process, resulting in highly super-Eddington mass accretion rates, accompanied by an optically-thick outflow emanating from the SMBH vicinity. This produces a week-month long, bright optical/Ultraviolet flare, accompanied by a year-decade long X-ray precursor and post-cursor emitted from the accretion flow onto the SMBH. Such ``Circular Tidal Disruption Events (TDEs)" represent a new class of nuclear transients, occurring at up to $1-10\%$ of the canonical parabolic tidal disruption event rate. Near breakup rotation and strong tidal deformation of the star prior to disruption could lead to strong magnetic fields, making circular-TDEs possible progenitors of jetted TDEs. Outflows prior to the final stellar disruption produce a circum-nuclear environment (CNM) with $\sim \rm 10^{-2} \, M_\odot$ at distances of $\sim 0.01-0.1 \, \rm pc$, likely leading to bright radio emission, and also similar to the CNM inferred for jetted TDEs. We discuss broader connections between circular TDEs and other recently identified classes of transients associated with galactic nuclei, such as repeating-TDEs and Quasi-Periodic X-ray Eruptions, as well as possible connections to luminous fast blue optical transients such as AT2018cow. We also discuss observational signatures of the analogous RLO of a white dwarf around an intermediate mass BH, which may be a multi-messenger source in the LISA era.
\end{abstract}


\section{Introduction} \label{sec:intro}

When a star happens to pass too close to a supermassive black hole (SMBH) residing in the center of a galaxy, it is disintegrated by the SMBH's tidal field, producing a bright multi-band electromagnetic transient. This process, known as a Tidal Disruption Event (TDE), was first studied by theorists starting in the 1970's \citep[e.g.,][]{Hills_1975,Lidskii_1979,Rees_1988,Evans_Kochanek_1989}, and became a rapidly growing observational field during the last three decades, with over 100 candidates to date \citep{Gezari_2021,Sazanov_21,Hammerstein_23,Yao_23,Masterson_24}. Many thousands of additional TDEs are expected to be discovered in the coming decade, with upcoming wide-field surveys such as the Rubin Observatory (LSST), ULTRASAT, UVEX and the Roman space telescope. 

The canonical theoretical picture regarding the formation of TDEs is as follows: Angular momentum relaxation within the nuclear star cluster  occasionally brings a star onto a catastrophic ``loss-cone'' orbit whose pericenter distance penetrates the tidal disruption radius, $\rtidal$. As the flux of stars scattered into the loss-cone is usually dominated by orbital separations comparable to the SMBH's radius of influence, $\rh\approx 1 \, \rm pc$, the disruption of the star thus occurs on a nearly radial orbit, with $1-e \approx \rtidal/\rh \lesssim 10^{-5}$ \citep[e.g.,][]{Stone_2020}. Other possible TDE formation channels include the capture of stars on retrograde orbits by a newly formed  gaseous AGN disk around the SMBH \citep{Wang+24}, or deviations from a spherically symmetric potential, resulting in ``centro-philic'' orbits that approach the SMBH at distances $\lesssim \rtidal$ \citep[e.g.,][]{Vasiliev_Merritt_2013}.

A small fraction ($\approx 10^{-3}-10^{-1}$) of the stars that feed the SMBH approach the tidal radius on mildly eccentric $(e \lesssim 0.5$), or even nearly circular ($e\ll 1$), orbits \citep{Linial_Sari_2017,Linial_Sari_2023,Metzger_2021,Lu_Quataert_23}, rather than on a parabolic orbit.  Many of these are likely formed by the disruption of stellar binaries by the tidal field of the SMBH, in which one star is ejected from the system at high velocities and one is bound to the BH \citep{Hills_1988}. As a star spirals in closer to the SMBH via the combined effects of gravitational wave (GW) emission and angular momentum diffusion, it may  begin to secularly shed mass onto the SMBH via Roche-lobe overflow. These GW-driven inspiralling stars, known as ``stellar-EMRIs'' (Extreme Mass Ratio Inspirals), have been invoked in the last few years as the origin of the emerging class of Quasi-Periodic Eruptions (QPEs) detected in X-rays \citep{Miniutti_2019,Giustini_2019,Arcodia_2021,Metzger_2021,Krolik_Linial_2022,Linial_Sari_2023,Lu_Quataert_23,Linial_Metzger_23}. 

In this paper we consider the electromagnetic transient that accompanies the unstable Roche-lobe overflow of a star on a nearly circular orbit around the SMBH, and its eventual disruption. While the mass transfer rate from the star to the SMBH is initially modest, it gradually accelerates to highly super-Eddington rates in the final months before before the final disruption of the star, producing an extended optically thick, super-Eddington outflow that enshrouds the system. Following the star's ultimate disruption, the remaining stellar material feeds the SMBH through a viscously spreading accretion disk. These "circular-TDEs" represent a unique and rare class of nuclear transients, which as we show, are characterized by years-decades episodes of bright soft X-ray emission, and a blue optical/UV flare lasting for days-month. Rapid stellar rotation induced by tides could lead to strong magnetic fields, making circular-TDEs potential sites for the formation of relativistic jets, while strong outflows prior to the complete disruption of the star may lead to bright accompanying radio emission. We further consider the case of a white-dwarf (WD) undergoing Roche-lobe overflow towards an Intermediate Mass Black Hole (IMBH), and discuss the multi-messenger detection prospects of such a system in the era of space-based GW detectors era.

The paper is organized is as follows. We begin by discussing the inspiral of the star and subsequent mass-trasnfer in section \ref{sec:Inspiral_Mass_Transfer}, followed by a generic calculation of the emission from a super-Eddington outflow engulfing the star-SMBH system (section \ref{sec:EM_from_wind}). We consider the different evolutionary phases of mass transfer and accretion in section \ref{sec:Temporal_Evolution}, and calculate approximate bolometric and band-dependent lightcurves in section \ref{sec:Lightcurves}. We discuss formation channels and rates in section \ref{sec:Formation}, and consider the analogous scenario of a WD and an IMBH in section \ref{sec:WD_Circular_TDE}, with conclusions and further discussion in section \ref{sec:Discussion}. Appendix \ref{sec:AppendixTidalHeating} summarizes results concerning the tidal heating of stars and WDs on circular and nearly circular orbits, near the onset of mass transfer.

\section{Inspiral and Mass Transfer} \label{sec:Inspiral_Mass_Transfer}

Consider a star of mass $m_\star$ and radius $R_\star$ orbiting a supermassive black hole (SMBH) of mass $\MBH$, where $m_\star \ll \MBH$. Assuming a circular orbit of radius $a$, the orbit decays due to GW emission on a timescale \citep[e.g.,][]{Peters_1964}
\begin{equation} \label{eq:t_gw_gen}
    \tgw \approx \frac{a}{|\dot{a}|} \approx \frac{5}{64} \frac{R_{\rm g}}{c} \pfrac{a}{R_{\rm g}}^4 \pfrac{\MBH}{m_\star} \,,
\end{equation}
where $R_{\rm g} = G\MBH/c^2$ is the SMBH's gravitational radius.

Mass transfer ensues once the orbital separation is of order the star's tidal radius, $\rtidal = R_\star (\MBH/m_\star)^{1/3}$. We define the separation at which Roche lobe overflow (RLO) first occurs as $r_{\rm MT} = \amt \rtidal$, with a corresponding orbital period
\begin{equation}
    P_{\rm orb} = 2\pi \amt^{3/2} \tdyn \,,
\end{equation}
that is notably independent of $\MBH$, and where $\tdyn = \sqrt{R_\star^3/Gm_\star}$ is the secondary's dynamical time. In the absence of tidal heating of the orbiting star, we expect the onset of RLO at $\amt \simeq 2$ for the extreme mass ratios considered in this paper \citep[e.g.,][]{Eggleton_83}. We show in Appendix \ref{sec:AppendixTidalHeating}, however, that except for very circular orbits with eccentricities $e \lesssim 10^{-2}-10^{-3}$ tidal heating is likely to inflate the radius of the star once $a \simeq (4-5)\times \rtidal$ leading to the onset of RLO at $\amt \simeq 4-5$, prior to when it would have occurred, in the absence of tidal heating.

In what follows, we shall focus on a main-sequence (MS) star of mass $m_\star \lesssim \rm M_\odot$. A related scenario, of a low-mass white dwarf (WD) undergoing mass transfer to an IMBH is discussed in \S \ref{sec:WD_Circular_TDE}.

Assuming a mass-radius relation of $R_\star\propto m_\star^{0.8}$ \citep{KW_1994} appropriate for MS stars (neglecting modifications due to tidal heating), the orbital period at the onset of mass transfer is
\begin{equation} \label{eq:PtidalMS}
    P_{\rm orb} \approx \\
    22 \, {\rm hr} \; \pfrac{\amt}{4}^{3/2} \pfrac{m_\star}{\rm M_\odot}^{0.7} \,,
\end{equation}
well outside the black hole's horizon and innermost stable circular orbit
\begin{equation} \label{eq:r_MT_over_Rg_MS}
    \frac{r_{\rm MT}}{R_{\rm g}} \approx 188 \, \pfrac{m_\star}{\rm M_\odot}^{0.47} M_{\bullet,6}^{-2/3} \pfrac{\amt}{4} \,,
\end{equation}
where $M_{\bullet,\rm X} = M_{\bullet}/(10^{\rm X} \, \rm M_\odot)$.

The GW inspiral timescale at this separation is roughly (Eq.~\ref{eq:t_gw_gen}, and see also \citealt{Dai_Blandford_2013,Linial_Sari_2017})
\begin{equation} \label{eq:t_gw_ms}
    \tgw \approx 10^7 \, {\rm yr} \; M_{\bullet,6}^{-2/3} \pfrac{m_\star}{\rm M_\odot}^{0.87} \pfrac{\amt}{4}^4 \,.
\end{equation}

\subsection{Mass Transfer}
The subsequent evolution of the system greatly depends on the mass transfer stability - whether it saturates to a self-regulating rate, or whether it accelerates in a runaway process.

If the resulting mass transfer is stable, its rate is set by angular momentum dissipation, such that the equilibrium $\dot{m}_\star$ is approximately
\begin{multline} \label{eq:Mdot_eq_MS}
    \dot{\mathcal{M}}_{\rm eq,GW} \approx \frac{m_\star/\tgw}{\dot{M}_{\rm Edd}}\approx \\
    2\times 10^{-6} \; \epsilon_{-1} \, M_{\bullet,6}^{-1/3} \pfrac{m_\star}{\rm M_\odot}^{0.13} \pfrac{\amt}{4}^{-4} \; \rm (MS) \,.
\end{multline}
where $\dot{M}_{\rm Edd} \equiv 4\pi G\MBH/(\kappa c \epsilon)$ is the Eddington accretion rate, $\kappa \approx 0.34 \, \rm cm^2 \, g^{-1}$ is the electron-scattering opacity and $\epsilon \equiv 0.1 \epsilon_{-1}$ is a fiducial radiative efficiency (not necessarily the true radiative efficiency). The relatively long GW inspiral time of a main-sequence star at $r_{\rm MT}$ thus results in a sub-Eddington mass transfer rate, provided that it is stable \citep[e.g.,][]{Dai_Blandford_2013,Linial_Sari_2017}\footnote{If indeed tidal heating is important, the onset and driving of RLO is dictated by the tidal heating rate rather than the GW inspiral time. In that case, $\dot{m}_\star \approx m_\star/(E_\star/\dot{E}_{\rm tide})$, where $E_\star$ is roughly the star's binding/internal energy, and $\dot{E}_{\rm tide}$ is the star's tidal heating rate - an extremely sensitive function of the pericenter distance and the orbital eccentricity. As discussed in Yao \& Quataert (in prep), this can in principle lead to much higher stable mass transfer rates than due to GW orbital decay alone.  We focus on unstable mass transfer in this paper and our results are not sensitive to significant changes in the equilibrium stable mass-transfer rate.}.

However, the proximity to the SMBH, the high mass ratio, the secondary's radial response to mass loss and the substantial mass loss through L2 (a consequence of the extreme mass ratio, $m_\star \ll \MBH$) suggest that mass transfer is highly non-conservative for many stars (particularly those that are fully convective), and is therefore likely to be \textit{unstable} \citep[see further discussion in e.g.,][]{Linial_Sari_2017,Linial_Sari_2023,Lu_Quataert_23}. In this case, $\dot{\mathcal{M}}_\star \equiv \dot{m}_\star/\dot{M}_{\rm Edd}$ increases in a runaway process, up to the very final disruption of the star, where Eddington rates as high as $\dot{\mathcal{M}}_\star \lesssim (m_\star/\tdyn)/\dot{M}_{\rm Edd} \approx 10^{6}$ are achieved\footnote{\cite{Lu_Quataert_23} proposed that in the case of a main-sequence star, the runaway evolution is terminated due to interaction between the star and the accretion flow it produces, resulting in an equilibrium mass accretion rate that never approaches $\dot{\mathcal{M}}_\bullet \gg 1$.   Whether this equilibrium can indeed be realized is uncertain, however, and depends sensitively on the uncertain structure of the accretion disk formed by stellar mass transfer. Here we consider the opposite limit in which the runaway mass transfer cannot be halted through star-disk interactions.}. In this paper we focus on this regime of unstable mass transfer and assess its observational consequences.  We note, however, that for inspiraling stars with finite eccentricity, the mass transfer stability is also likely  sensitive to the structure of the tidally heated star, which can be very different from a normal MS star or WD; this will be considered in future work.

The density stratification in the outer layers of the secondary suggests that the rate of mass transfer sensitively depends on the extent to which the star overfills its Roche lobe. We define
\begin{equation} \label{eq:xi_def}
    \xi=(R_\star-R_{\rm RL})/R_\star \,,
\end{equation}
where $R_{\rm RL} \approx \frac{1}{2} a (m_\star/\MBH)^{1/3}$ is the Roche lobe radius \citep{Eggleton_83}. Assuming that the outer layers of the star are well-described as a polytrope, the mass loss rate can be approximated
\citep[e.g.,][]{Ritter_1988,Linial_Sari_2017,Linial_Sari_2023}
\begin{equation} \label{eq:Mdot_xi_gen}
    \dot{m}_\star \approx \frac{m_\star}{\tdyn} \xi^k \,,
\end{equation}
where $k\approx 3$ for a polytropic index of $3/2$, appropriate for the low-mass MS stars considered here. {The value of $k$ may be different in tidally heated stars (Appendix \ref{sec:AppendixTidalHeating}), but we focus on $k \simeq 3$ for concreteness throughout this paper.} 

Unstable mass transfer proceeds when the growth of $\xi$ is dominated by the stellar expansion following the stripping of mass, rather than by the orbital decay due to GWs (or, the tidal heating timescale, if tides are responsible for driving mass transfer, see footnote 1). The system's evolution time, or the time spent at accretion rate $\dot{m}_\star$ is then given by
\begin{equation} \label{eq:tau_mdot}
    \frac{\dot{m}}{\ddot{m}} \approx \tau_{\dot{m}} \approx \tau_m \pfrac{\tdyn}{\tau_m}^{1/k} \,,
\end{equation}
where $\tau_m = m_\star/\dot{m}_\star$, up to some order unity factor that will depend on the details of the stellar structure and the degree of angular momentum conservation (or lack of). Note that $\tau_{\dot m} < \tau_{m}$ up to the final disruption of the star, namely, the time spent at around accretion rate $\dot{m}_\star$ interval is generally shorter than $m_\star/|\dot{m}_\star|$.

\section{The emission from an optically thick, Super-Eddington outflow} \label{sec:EM_from_wind}

{The loss of mass from the secondary injects mass at a rate $\dot{m}_\star$ at radius $r_{\rm MT}$, initially possessing specific angular momentum similar to that of the secondary, $\sqrt{G\MBH r_{\rm MT}}$, forming circular tori of gas near the L1 and L2 Lagrange points \citep[e.g.,][]{Shu_Lubow_Anderson_79,Hubova_Pejcha_19}. The development of turbulence through the magneto-rotational instability (MRI) leads to the transport of angular momentum and accretion of mass towards the SMBH at rate $\dot{m}_\bullet$. A similar setup was studied in 3-dimensional global radiation magentohydrodynamical simulations by \cite{Jiang_2019}, showing that at the super-Eddington rates we are interested in, quasi-spherical, radiation-dominated outflows of velocity $v_{\rm w} = \beta_{\rm w} c$ are launched with $\beta_{\rm w} \approx 0.1$, carrying a significant fraction $f_{\rm w} \lesssim 1$ of the total injected mass, $\dot{m}_{\rm w} = f_{\rm w} \dot{m}_\bullet$.} In general, $\dot{m}_\bullet$ does not necessarily trace the mass loss rate from the star.  However, if a steady-state is established, i.e.,  when $\dot{m}_\star$ evolves on timescales longer than the viscous evolution time, we have $\dot{m}_\bullet \approx \dot{m}_\star / (1+f_{\rm w})$, such that $\dot{m}_\star=\dot{m}_\bullet+\dot{m}_{\rm w}$ (excluding secular changes to the disk mass).

{In what follows, we consider an idealized form of the resulting outflow to allow for analytical estimates of the electromagnetic emission associated with the runaway evolution of a companion undergoing RLO near an SMBH. We make the following simplifying assumptions: (a) the outflow is quasi-steady and traces the accretion rate onto the SMBH, $\dot{m}_\bullet$, (b) the outflow is spherical, propagating as a wind of a fixed velocity $v_{\rm w}$, (c) $f_{\rm w}$ and $\beta_{\rm w}$ are independent of $\dot{\mathcal{M}}_\bullet \equiv \dot{m}_\bullet/\dot{M}_{\rm Edd} \gg 1$ and $\MBH$, (d) the outflow is launched from the vicinity of the SMBH, at roughly $R_{\rm g} \beta_{\rm w}^{-2}$.}

The density of the assumed spherically symmetric wind-like outflow is given by
\begin{equation}
    \rho(r) = \frac{\dot{m}_{\rm w}}{4\pi v_{\rm w} r^2} \,,
\end{equation}
such that the optical depth from radius $r$ to infinity is
\begin{equation}
    \tau(r) = \int_r^{\infty} \kappa \rho(r') \, d r' \approx \frac{\kappa \dot{m}_{\rm w}}{4\pi v_{\rm w} r} \,,
\end{equation}
for constant opacity $\kappa$.

Radiation is trapped in the outflow as long as $\tau(r) \gg c/v_{\rm w}$, or interior to the trapping radius
\begin{equation}
    r_{\rm tr} \approx \frac{\kappa \dot{m}_{\rm w}}{4\pi c} = R_{\rm g} f_{\rm w} \epsilon^{-1} \dot{\mathcal{M}}_\bullet \,,
\end{equation}
where we assumed that opacity is dominated by electron scattering.

Up to $r\lesssim r_{\rm tr}$, the flow is nearly adiabatic and radiation-dominated, hence $p_{\rm rad} \propto \rho^{4/3} \propto r^{-8/3}$. The radiative flux at the trapping radius is given by
\begin{equation}
    L \approx \frac{4\pi r_{\rm tr}^2 p_{\rm rad}(r_{\rm tr}) c}{\tau(r_{\rm tr})} \approx 4\pi r_{\rm tr}^2 v_{\rm w} p_{\rm rad}(r_{\rm s}) \pfrac{r_{\rm tr}}{r_{\rm s}}^{-8/3} \,,
\end{equation}
where $r_{\rm s} \approx R_{\rm g} \beta_{\rm w}^{-2}$ is the sonic radius at which is the outflow is being launched. Given the assumed $\beta_{\rm w}\approx 0.1$, the launching radius is of order $r_{\rm MT}$ (Eq.~\ref{eq:r_MT_over_Rg_MS}). At this radius, $p_{\rm rad}(r_{\rm s}) \approx \rho(r_{\rm s}) v_{\rm w}^2$, and thus the luminosity can be expressed as
\begin{equation}
    L \approx \dot{m}_{\rm w} v_{\rm w}^2 \pfrac{r_{\rm tr}}{r_{\rm s}}^{-2/3}\,,
\end{equation}
or
\begin{equation}
    L \approx L_{\rm Edd} \pfrac{\dot{E}_{\rm w}}{L_{\rm Edd}}^{1/3} \,,
    \label{eq:LsuperEdd}
\end{equation}
where $\dot{E}_{\rm w} \equiv \dot{m}_{\rm w} v_{\rm w}^2 =  L_{\rm Edd} \dot{\mathcal{M}}_\bullet \epsilon^{-1} f_{\rm w} \beta_{\rm w}^2$ is the kinetic power carried by the wind (up to a factor of 2); this is similar to the results of \cite{Strubbe_Quataert_2009}. In terms of Eddington mass accretion ratio, the luminosity can be expressed as
\begin{equation} \label{eq:L_Mdot}
    L\approx L_{\rm Edd} \dot{\mathcal{M}}_\bullet^{1/3} \epsilon^{-1/3} f_{\rm w}^{1/3} \beta_{\rm w}^{2/3} \,.
\end{equation}
which may greatly exceed $L_{\rm Edd}$ as $\dot{\mathcal{M}}_\bullet \gg 1$. We assume that $f_{\rm w}$ and $\beta_{\rm w}$ are nearly independent of $\dot{\mathcal{M}}_\bullet$ (assumption (c)), such that the luminosity may vary substantially during the runaway phase, as $\dot{\mathcal{M}}_\bullet$ increases by several order of magnitude. Equations \ref{eq:LsuperEdd} \& \ref{eq:L_Mdot} show that the simplest model for the radiated luminosity of  super-Eddington outflows predicts that most of the power supplied to the wind at the base ends up in kinetic energy, not radiation.   If, however, there are internal processes that dissipate the wind kinetic energy at larger radii (e.g., `internal shocks,' shear between gas with different speeds at different polar angles), this will increase the radiated luminosity relative to equation \ref{eq:L_Mdot}.  In this sense, our predictions that follow may underestimate the luminosity during the super-Eddington outflow phase.

Since the outflow velocity $v_{\rm w}$ is not much smaller than $c$, the trapping radius and the photosphere are not too different, and we thus evaluate the emission's black-body temperature at $r_{\rm tr}$
\begin{multline} \label{eq:T_BB_wind}
    \kb T_{\rm BB} \approx \kb \pfrac{L\, c/v_{\rm w}}{4\pi r_{\rm tr}^2 
  \sigma_{\rm SB}}^{1/4} \approx \\
  \kb \pfrac{c^5 }{G \MBH \kappa \sigma_{\rm SB} }^{1/4} \dot{\mathcal{M}}_\bullet^{-5/12} (\epsilon/f_{\rm w})^{5/12} \beta_{\rm w}^{-1/12} = \\
  150 \, {\rm eV} \; M_{\bullet,6}^{-1/4} \dot{\mathcal{M}}_\bullet^{-5/12} (\epsilon/f_{\rm w})^{5/12} \beta_{\rm w}^{-1/12} \,,
\end{multline}
where $\sigma_{\rm SB}$ is the Stefan-Boltzmann constant. The temperature decreases with increasing luminosity as \begin{multline} \label{eq:L_T_obs}
    L(T_{\rm BB}) \approx 4.4\times 10^{44} \, {\rm erg \, s^{-1}} \,
    \pfrac{\kb T_{\rm BB}}{100 \, \rm eV}^{-4/5} M_{\bullet,6}^{4/5} \, \beta_{\rm w}^{3/5} \,.
\end{multline}
where we assumed that efficient thermalization is achieved in the outflow.

We note that our assumption concerning the sphericity of the outflow is inaccurate in light of the strong dependence of the flow properties on the polar angle, as found in \cite{Jiang_2019}, suggestive of a conical (i.e., hourglass shaped), rather than spherical, outflow. Photon diffusion in the lateral direction may dominate the cooling of the ejecta in the case of a narrow conical outflow.   However, since the typical opening angle of the outflow obtained in \cite{Jiang_2019} is of order unity, we expect this to result in mild corrections to our estimates.

\begin{figure*}
    \centering
    \includegraphics[width=\textwidth]{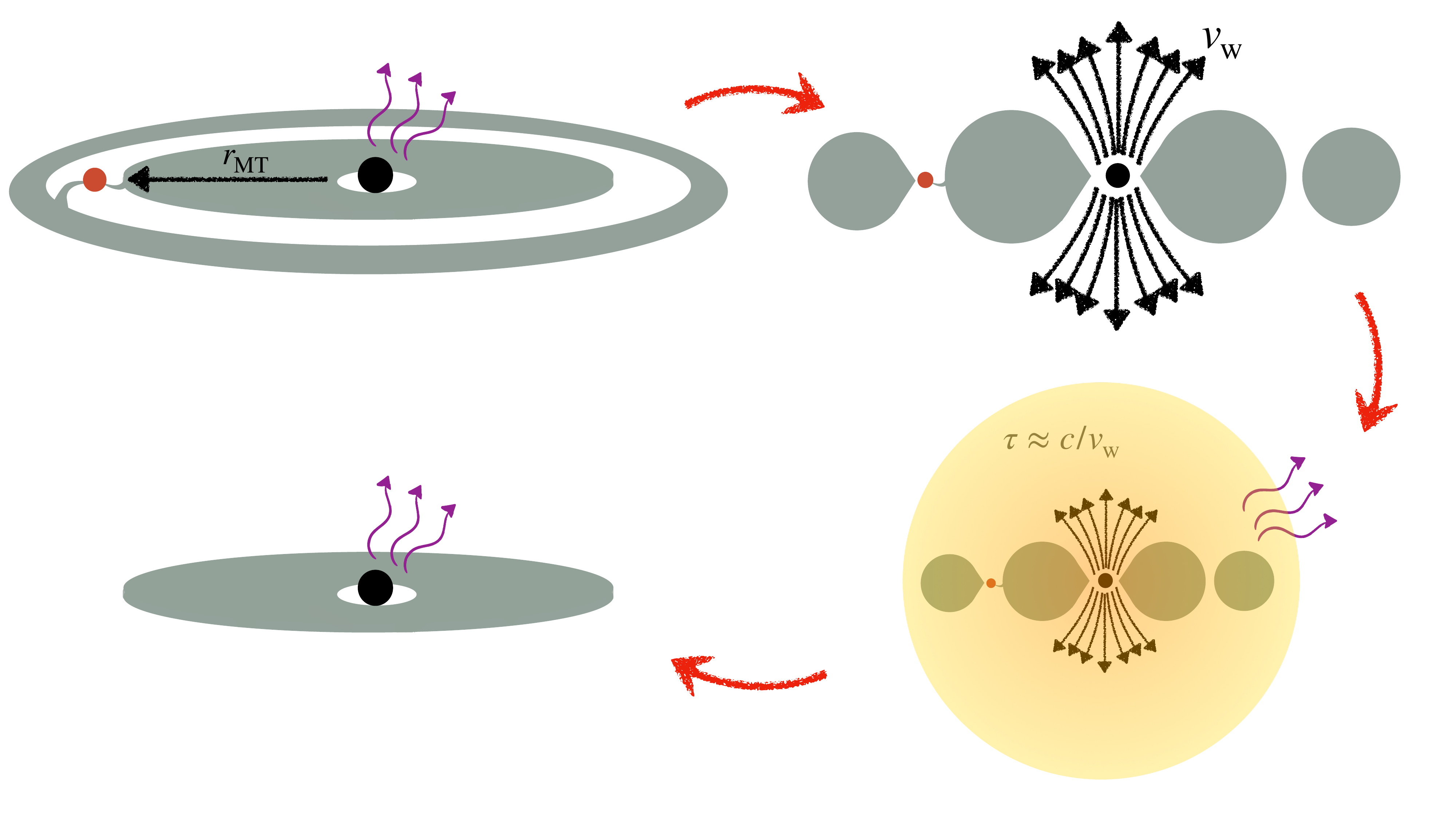}
    \caption{A schematic sketch indicating the different stages of evolution. Mass transfer from the star to the SMBH ensues once $r = r_{\rm MT} \approx {\rm few} \times \rtidal$. At low accretion rates ($\dot{m}_\star \lesssim \dot{M}_{\rm Edd}$) matter leaving the star from the inner Lagrange point produces a thin accretion disk, emitting X-rays (alongside a circumbinary ring fed by the outer Lagrange point). Mass transfer continues to accelerate to super-Eddington rates, producing quasi-spherical, optically-thick outflows. The extended photosphere produces super-Eddington emission peaking at optical/UV frequencies. The star is eventually fully disrupted, with its remnants feeding the SMBH through viscous evolution, eventually emitting X-rays from a radiatively efficient thin disk.}
    \label{fig:Cartoon}
\end{figure*}

\section{Temporal evolution}
\label{sec:Temporal_Evolution}

In the previous section we discussed the emission associated with an SMBH fed at a highly super-Eddington rate, $\dot{\mathcal{M}}_\bullet \gg 1$. Here we describe the accretion rate onto the SMBH as a function of time, and derive approximate lightcurves. We begin by discussing three distinct temporal phases, over which $\dot{\mathcal{M}}_\bullet$ varies by several orders of magnitude, also shown in Fig.~\ref{fig:Cartoon}: an initial mass transfer driven by GW inspiral, a phase of runaway unstable mass transfer and finally, a late time accretion phase onto the SMBH set by viscous evolution of the tidally stripped debris of the victim star.

\subsection{Onset of mass transfer and GW driven phase}
As $a$ approaches $r_{\rm MT}$, mass transfer ensues, initially driven by GW inspiral, bringing the secondary to progressively overfill its Roche lobe at a rate $|\dot{\xi}| \approx \tgw^{-1}$. The resulting mass transfer therefore initially increases with time as 
\begin{equation} \label{eq:Mdot_pre_runaway}
    \dot{m}_\star \approx \frac{m_\star}{\tdyn} \pfrac{t}{\tgw}^{k} \,,
\end{equation}
where $t=0$ marks the onset of mass transfer. {When in the sub-Eddington regime of $\dot{\mathcal{M}}_\bullet \lesssim 0.01$, a radiatively inefficient accretion flow develops, with a radiative efficiency likely $\lesssim 1 \%$ (e.g., \citealt{Sharma2007,Yuan2014}).   We do not model the radiation from this phase in this paper but instead focus on the brighter, shorter-lived high accretion rate phases detectable in time-domain surveys.  The long-lived low-Eddington phase could, however, contribute to the luminosity function of low-luminosity active galactic nuclei in nearby galaxies.}

\subsection{Runaway evolution}

The runaway discussed in \S \ref{sec:Inspiral_Mass_Transfer} dominates the evolution once the mass transfer rate predicted by Eq.~\ref{eq:Mdot_pre_runaway} exceeds the equilibrium value associated with stable mass transfer, $m_\star/\tgw$ (Eq.~\ref{eq:Mdot_eq_MS}). The transition into the runaway phase therefore occurs at time\footnote{The onset of runaway mass-transfer may occur somewhat closer to disruption (smaller $t_{\rm runaway}$) if stable mass transfer is set by tidal heating, but this does not significantly modify the unstable mass-transfer phases that are the focus of this paper.}
\begin{equation}
    t_{\rm runaway} \approx \tgw (\tdyn/\tgw)^{1/k} \,,
\end{equation}
and for a MS star
\begin{equation} \label{eq:tau_runaway_MS}
    t_{\rm runaway} \approx 2400 \, {\rm yr} \; M_{\bullet,6}^{-4/9} \pfrac{m_\star}{\rm M_\odot}^{0.8} \pfrac{\amt}{4}^{8/3} \,.
\end{equation}
Physically, this is the time at which Roche lobe overflow proceeds primarily due to the radial response of the star as a result of mass loss, rather than the GW inspiral which brought it to overfill its Roche lobe in the first place. Mass transfer then evolves on ever decreasing timescales, with $\dot{m}_\star \propto (t_f - t)^{-1/(1-1/k)} \approx (t_f - t )^{-1.5}$, where $t_f \approx 2 t_{\rm runaway}$ is the time at which the secondary is fully disrupted (i.e., $\xi \approx 1$). 

When the accretion rate increases to the range of $0.01 \lesssim \dot{\mathcal{M}}_\bullet \lesssim 1$, a radiatively efficient, geometrically thin accretion disk is formed, radiating at a bolometric luminosity $L_{\rm bol} \approx \dot{\mathcal{M}}_\bullet L_{\rm Edd} = \epsilon \,\dot{m}_\bullet c^2$, and a characteristic blackbody temperature set by the disk's inner radius, $R_{\rm in} \approx 4 R_{\rm g}$, at roughly 
\begin{multline} \label{eq:T_BB_disk}
    \kb T_{\rm BB,disk} \approx k_{\rm BB} \pfrac{3G\MBH \dot{m}_\bullet}{8\pi \sigma_{\rm SB} R_{\rm in}^3}^{1/4} \approx \\
    100 \, {\rm eV} \, \epsilon_{-1}^{-1/4} \, \dot{\mathcal{M}}_\bullet^{1/4} M_{\bullet,6}^{-1/4} \,,
\end{multline}
where $\sigma_{\rm SB}$ is the Stefan-Boltzmann constant. This radiatively efficient phase commences when $\dot{\mathcal{M}}_\bullet \approx 0.01$, at time
\begin{equation} \label{eq:transition_to_1pEdd_MS}
    t_f-t_{1\% \, \rm Edd} \approx 9 \, {\rm yr} \; \epsilon_{-1}^{2/3} \, M_{\bullet,6}^{-2/3} \pfrac{m_\star}{\rm M_\odot}^{0.9} \,.
\end{equation}
while an Eddington accretion rate $\dot{\mathcal{M}}_\bullet \approx 1$ is first achieved at time
\begin{equation} \label{eq:transition_to_Edd_MS}
    t_f-t_{\rm Edd} \approx 0.4 \, {\rm yr} \; \epsilon_{-1}^{2/3} M_{\bullet,6}^{-2/3} \pfrac{m_\star}{\rm M_\odot}^{0.9} \,,
\end{equation}
before the final disruption of the star. The system thus spends multiple years as a bright source of soft X-rays, with $0.01 \lesssim L_{\rm bol}/L_{\rm Edd} \lesssim 1$. {There is also an extended (but somewhat shorter) phase above $\dot{\mathcal{M}}_\bullet \gtrsim 0.01$ after the star's disruption, in the viscously dominated accretion phase discussed in \S \ref{sec:visc} below.}

Once $\dot{\mathcal{M}}_\bullet \gtrsim 1$, the emission will continue to evolve as described in \S \ref{sec:EM_from_wind}, being dominated by an optically thick, super-Eddington outflow. Assuming that $\dot{m}_\bullet \propto \dot{m}_\star$, the system's bolometric luminosity will increase in time as $L_{\rm bol} \propto \dot{\mathcal{M}}_\bullet^{1/3} \propto (t_f-t)^{-0.5}$ (Eq.~\ref{eq:L_Mdot}), and the corresponding temperature will decrease, with the system gradually brightening in softer bands (a result of the expanding trapping radius/photosphere).

\subsection{Viscously dominated accretion rate}
\label{sec:visc}
At late times, the supply of mass towards the SMBH becomes limited by the finite viscous time at radius $r_{\rm MT}$
\begin{multline} \label{eq:t_visc}
     t_{\rm visc} \approx \frac{P_{\rm orb}}{2\pi} \pfrac{h}{r_{\rm MT}}^{-2} \alpha^{-1} \approx \\
    16 \, {\rm d} \, \pfrac{h/r_{\rm MT}}{0.3}^{-2} \alpha_{-1}^{-1} \pfrac{m_\star}{\rm M_\odot}^{0.7} \pfrac{\amt}{4}^{3/2} \,,
\end{multline}
where $h$ is the disk scale height at $r_{\rm MT}$ and $\alpha \equiv 0.1 \, \alpha_{-1}$ is the dimensionless viscosity parameter \citep{Shakura1976}. Given the large uncertainties in $h/r_{\rm MT}$, $\alpha$ and $\amt$, $t_{\rm visc}$ is constrained at best to within a factor of $\sim 10$.

Once $\tau_{\dot{m}} \lesssim t_{\rm visc}$, the assumption that the accretion onto the SMBH traces the star's mass loss rate (assumption (a)), clearly fails, as the viscously limited supply decouples $\dot{m}_\star$ and $\dot{m}_\bullet$. While the star's mass loss rate continues to increase, $\dot{m}_\bullet$ is limited by the disk's feeding rate
\begin{equation}
    \dot{m}_\bullet = \dot{M}_{\rm d} \approx 3\pi\Sigma(r_{\rm MT}) \nu \approx \frac{\pi r_{\rm MT}^2 \Sigma(r_{\rm MT})}{(t_{\rm visc}/3)}\,,
\end{equation}
where the effective viscosity, $\nu$, is given by
\begin{equation}
    \nu \approx \alpha \sqrt{G\MBH r_{\rm MT}} \pfrac{h}{r_{\rm MT}}^2 \,.
\end{equation}
Shortly after the transition to the viscously dominated evolution, the star undergoes complete disruption, and the surface density, $\Sigma(r_{\rm MT})$ is given by the remaining mass of the star (which is essentially the entirety of the initial stellar mass, as most of the mass loss occurs at the very last few orbits), spread over an annulus of radius and thickness $\sim r_{\rm MT}$, namely $\Sigma(r_{\rm MT})\approx m_\star/(\pi r_{\rm MT}^2)$, yielding an Eddington ratio of roughly
\begin{multline} \label{eq:Mdot_visc_max}
    \dot{\mathcal{M}}_{\rm visc,0} \approx \frac{m_\star}{t_{\rm visc} \dot{M}_{\rm Edd}} \approx \frac{\alpha}{4\pi \amt^{3/2}} \frac{m_\star}{\MBH} \pfrac{h}{r_{\rm MT}}^2 \frac{c \kappa \epsilon}{G \tdyn} \approx \\
    9\times 10^2 \, \frac{\epsilon_{-1} \alpha_{-1}}{M_{\bullet,6}} \pfrac{\amt}{4}^{-3/2} \pfrac{h/r}{0.3}^2 \pfrac{m_\star}{\rm M_\odot}^{0.3} \,.
\end{multline}
The subsequent evolution is akin to a viscously spreading localized ring (i.e., a delta function in surface density), transporting angular momentum outward while feeding mass towards the SMBH. \cite{Metzger_2008} obtained self-similar solutions for the time-dependent evolution of an advective disk undergoing substantial viscously-driven outflows, accreting at super-Eddington rates, as is appropriate here. The feeding rate onto the SMBH, which also sets the outflow mass flux is given by \citep[appendix B of][]{Metzger_2008}
\begin{equation} \label{eq:Mdot_post_disruption}
    \dot{\mathcal{M}}_\bullet \approx \dot{\mathcal{M}}_{\rm visc,0} \pfrac{t - t_{f}}{t_{\rm visc}}^{-8/3} \,,
\end{equation}
{The exact power-law decay of the accretion rate after the star's disruption is somewhat uncertain.  For example,  
in the absence of substantial outflows, it would scale as $t^{-4/3}$ \citep{Metzger_2008}.  We take equation \ref{eq:Mdot_post_disruption} as a fiducial estimate of the late-time accretion rate but the temporal decay could be somewhat shallower depending on the exact strength of outflows from the disk and how much angular momentum they carry.}

A super-Eddington accretion rate proceeds for a duration of roughly
\begin{multline} \label{eq:transition_below_Edd_MS}
    \Delta t_{\rm Edd,\downarrow}\approx t_{\rm visc} \dot{\mathcal{M}}_{\rm visc,0}^{3/8} \approx 
    200 \, {\rm d} \, \\
    \pfrac{\epsilon_{-1}}{M_{\bullet,6}}^{3/8} \pfrac{h/r}{0.3}^{-5/4} \alpha_{-1}^{-5/8} \pfrac{\amt}{4}^{15/16} \pfrac{m_\star}{\rm M_\odot}^{0.8} \,, 
\end{multline}
after the final disruption of the star. Once the accretion rate becomes sub-Eddington, $\dot{\mathcal{M}}_\bullet$ no longer evolves according to Eq.~\ref{eq:Mdot_post_disruption}. Rather, the accretion flow settles again into a radiatively thin disk phase, akin to the regime discussed in \cite{Cannizzo1990} in the context of the asymptotic behavior of TDE accretion disks. Adopting their solution for a viscously spreading thin disk, we use $\dot{\mathcal{M}}_\bullet\propto t^{-1.2}$. $\dot{\mathcal{M}}_\bullet$ remains above $0.01$ for a duration of roughly $100^{1/1.2} \times \Delta t_{\rm Edd,\downarrow} \approx 20 \, \rm yr$.
Note that \cite{Cannizzo1990} have assumed that the effective viscosity is proportional to the midplane's gas pressure $\nu \propto p_{\rm gas}$, rather than the total, radiation and gas pressure, in order to circumvent the thermal and viscous instability of radiation pressure dominated alpha-disks. This ad-hoc simplification may not be well-justified in the inner parts of the disk when $\dot{\mathcal{M}}_\bullet$ is still close to 1.

\begin{figure}
    \centering
    \includegraphics[width=0.5\textwidth]{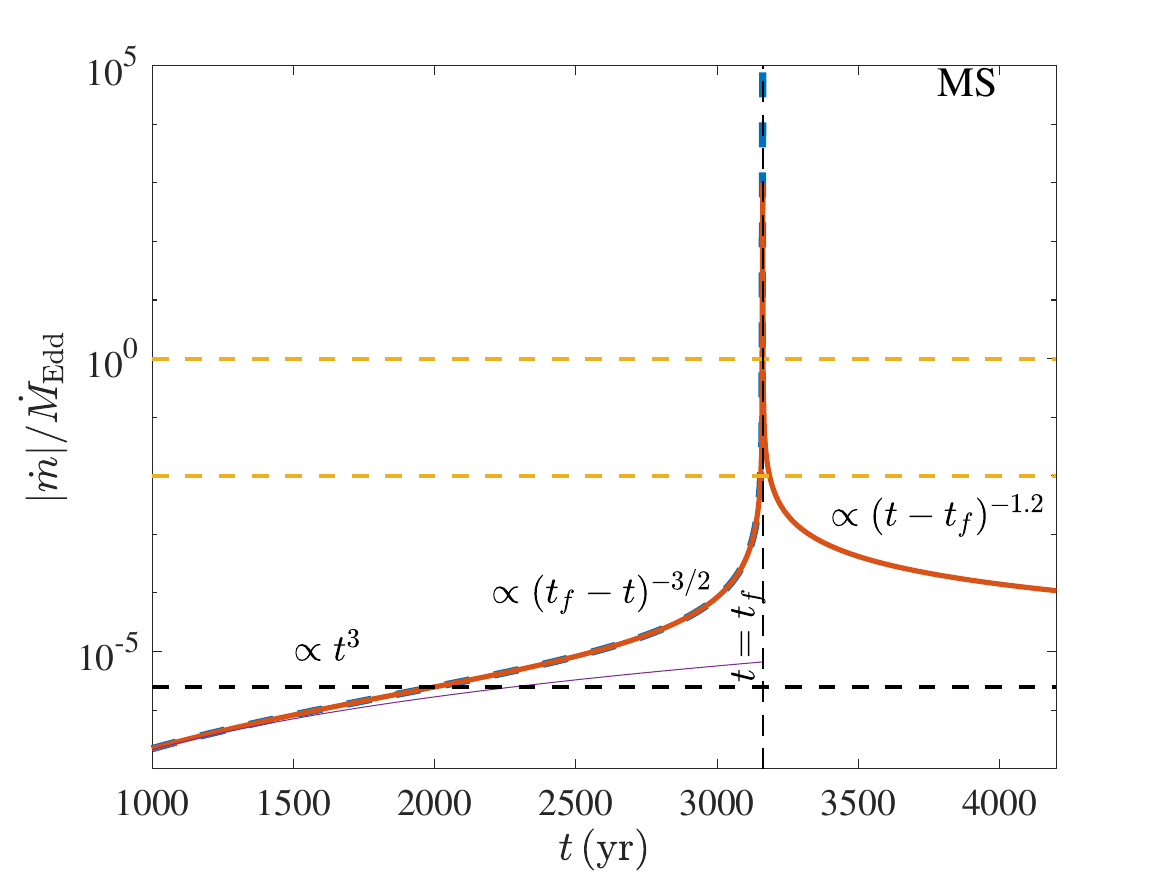}
    \caption{Temporal evolution of an unstable mass transfer from a Sun-like main-sequence star onto an SMBH of mass $\MBH=10^6 \, \rm M_\odot$. \textit{Dashed blue curve} is the mass loss rate $\dot{m}_\star$ in units of $\dot{M}_{\rm Edd}$ as a function of time $t$ since the onset of mass transfer. \textit{Red solid curve} is $\dot{m}_\bullet$, the feeding rate onto the SMBH, which mostly follows $\dot{m}_\star$, almost up to the final disruption of the star at time $t_f \approx 500 \, \rm yr$. The remnants of the star then accrete onto the SMBH through viscous spreading, decaying at times $t>t_f$. \textit{Horizontal black dashed line} is roughly the steady-state accretion rate through stable mass transfer (Eq.~\ref{eq:Mdot_eq_MS}), and the two \textit{horitzontal orange dahsed lines} represent Eddington ratios of 0.01 and 1, for which a radiatively efficient disc forms.}
    \label{fig:Mdot_vs_time}
\end{figure}

Figure \ref{fig:Mdot_vs_time} demonstrates the temporal evolution of the stellar mass loss rate and the corresponding accretion rate onto the SMBH, for our fiducial low-mass MS scenario. Here we solved the time evolution of the Roche-lobe overfilling, $\xi$,
\begin{equation}
    \dot{\xi} = (1-\xi) \left( \frac{2}{3 \tdyn} \xi^k + \frac{1}{\tgw} \right) \,,
\end{equation}
up until $\xi \gtrsim 0.5$, which we take as the star's final disruption, at which point the assumptions of the star's polytropic structure near its surface certainly become invalid. The above equation is obtained by taking the full time derivative of $\xi$ (using Eqs.~\ref{eq:xi_def}, \ref{eq:Mdot_xi_gen}), assuming that the star responds adiabatically to mass loss $R_\star \propto m_\star^{-1/3}$, and that mass transfer is completely non-conservative.

\section{Lightcurves} \label{sec:Lightcurves}
We now discuss the temporal evolution of the emission associated with the system. We consider the contribution of a radiatively efficient accretion disk, prevailing when $0.01 \lesssim \dot{\mathcal{M}}_\bullet \lesssim 1$, and the emission from an optically thick outflow produced when the accretion onto the SMBH is super-Eddington $\dot{\mathcal{M}}_\bullet \gg 1$.

\begin{figure}
    \centering
    \includegraphics[width=0.5\textwidth]{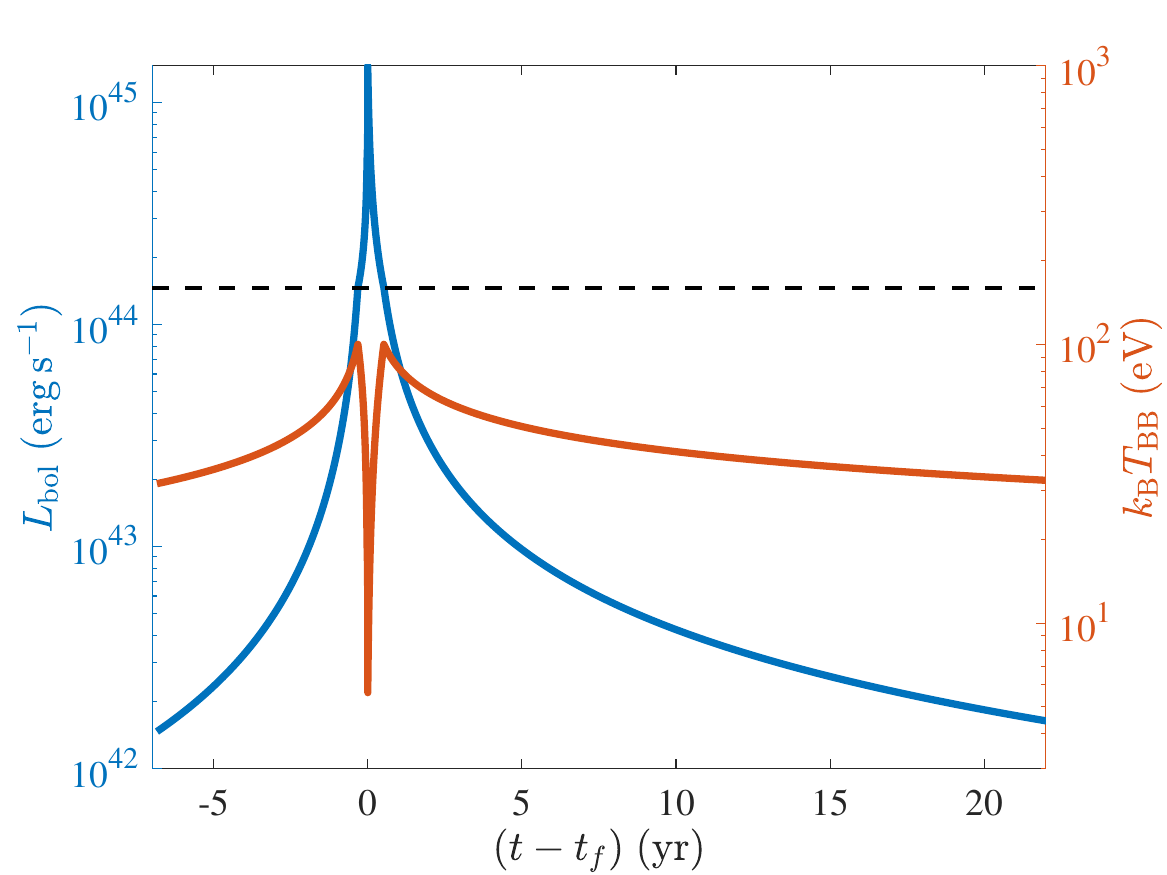}
    \caption{Bolometric luminosity and characteristic black-body temperature as a function of time for a Sun-like star orbiting a $10^6 \, \rm M_\odot$ SMBH. Time is measured relative to the star's final disruption, $t_{f}$. Horizontal \textit{black dashed line} is the SMBH's Eddington luminosity.}
    \label{fig:L_bol_T_BB_vs_t}
\end{figure}

Figure \ref{fig:L_bol_T_BB_vs_t} demonstrates the brightening of the emission up to the star's final disruption, followed by rapid dimming, governed by the remaining disk's viscous spreading. The characteristic black-body temperature initially increases as $T_{\rm BB} \propto \dot{\mathcal{M}}^{1/4}_\bullet \propto (t_f - t)^{-3/8}$, starting when $\dot{\mathcal{M}}_\bullet \lesssim 0.01$. At later times, when $\dot{\mathcal{M}}_\bullet \sim 1$, the emission temperature decreases as $T_{\rm BB} \propto \dot{\mathcal{M}}_\bullet^{-5/12} \propto (t_f - t)^{5/8}$, whereas in the viscously dominated evolution, at times $t>t_f$ the temperature initially increases as $T_{\rm BB} \propto \dot{\mathcal{M}}_\bullet^{-5/12} \propto (t - t_f)^{10/9}$, up until the accretion rate becomes sub-Eddington again, with the temperature falling off as $T_{\rm BB} \propto \dot{\mathcal{M}}^{1/4}_\bullet \propto (t_f - t)^{-2/3}$. To allow for continuous transition of temperature and luminosity between these different regimes, we include a correction factor in Eq.~\ref{eq:T_BB_wind}, such that it coincides with the inner disk temperature of Eq.~\ref{eq:T_BB_disk} at $\dot{\mathcal{M}}_\bullet$ (0.55 for $\beta_{\rm w} = f_{\rm w} = \epsilon = 0.1$).

Overall, the system's peak bolometric luminosity is given by (combining Eq.~\ref{eq:L_Mdot} and ~\ref{eq:Mdot_visc_max})
\begin{multline}    
    L_{\rm bol}^{\rm max} \approx
    3\times 10^{45} \; {\rm \ergss} f_{\rm w}^{1/3} \alpha_{-1}^{1/3} \beta_{\rm w}^{2/3} \\
    \pfrac{\amt}{4}^{-1/2}  \pfrac{h/r}{0.3}^{2/3} M_{\bullet,6}^{2/3} \pfrac{m_\star}{\rm M_\odot}^{0.1} \,,
\end{multline}
with a corresponding black-body temperature at peak luminosity is roughly
\begin{multline} \label{eq:kT_peak_MS}
    \kb T_{\rm BB} (L_{\rm bol}^{\rm max}) \approx 3.7 \, {\rm eV} \; f_{\rm w}^{-5/12} \alpha_{-1}^{-5/12} \beta_{\rm w}^{-1/12} \, M_{\bullet,6}^{1/6} \\
    \pfrac{\amt}{4}^{15/24} \pfrac{h/r}{0.3}^{-5/6}  \pfrac{m_\star}{\rm M_\odot}^{-0.12} \,,
\end{multline}
whose spectral energy-density peaks in the Extreme-UV (EUV), with $3k_{\rm B}T_{\rm BB} \gtrsim 10 \, \rm eV$.
The typical time spent around peak is comparable to the viscous time (Eq.~\ref{eq:t_visc}), such that the total emitted energy at around peak is approximately
\begin{multline}
    L_{\rm bol}^{\rm max} \times t_{\rm visc} \approx 4\times 10^{51} \, {\rm erg} \\
    f_{\rm w}^{1/3} \pfrac{\beta_{\rm w} M_{\bullet,6}}{\alpha_{-1}}^{2/3}
    \pfrac{h/r_{\rm MT}}{0.3}^{-4/3} \pfrac{m_\star}{\rm M_\odot}^{0.8} \pfrac{\amt}{4}.
\end{multline}

\begin{figure}
    \centering
    \includegraphics[width=0.5\textwidth]{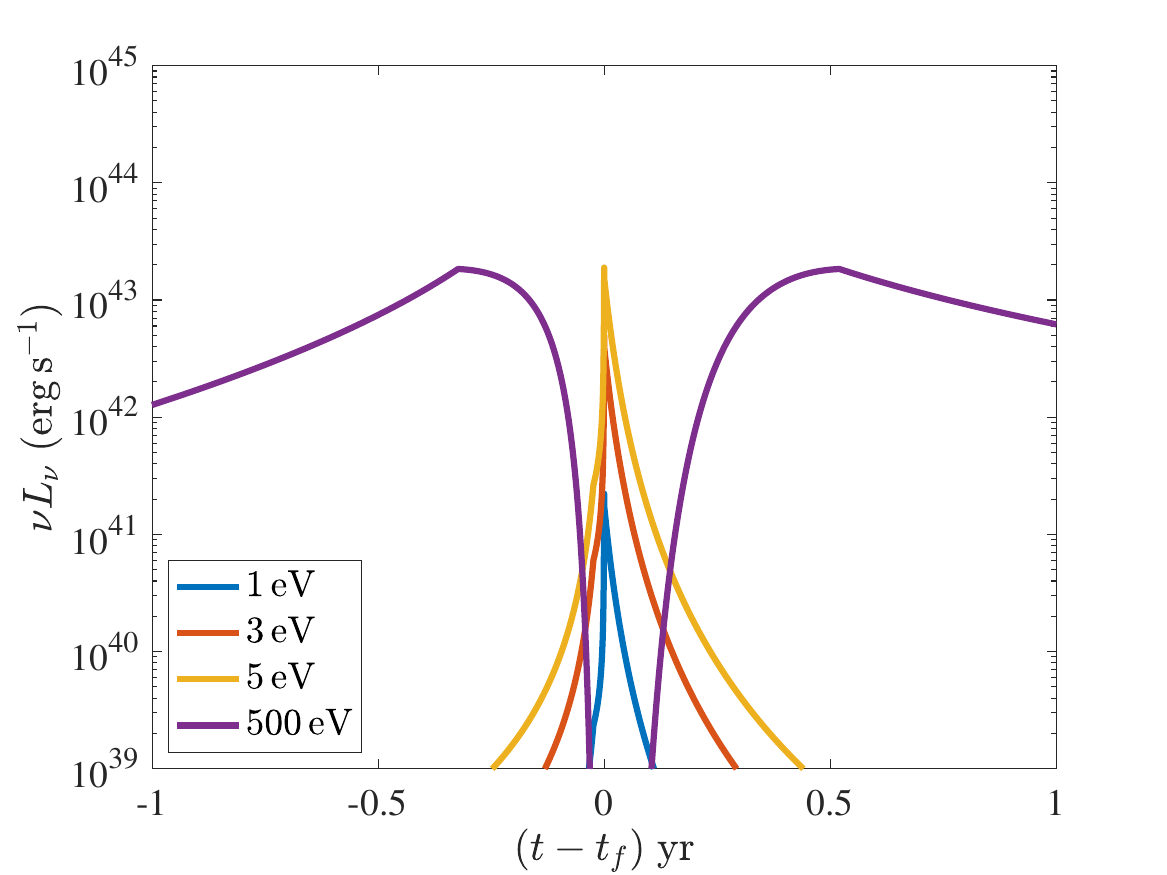}
    \includegraphics[width=0.5\textwidth]{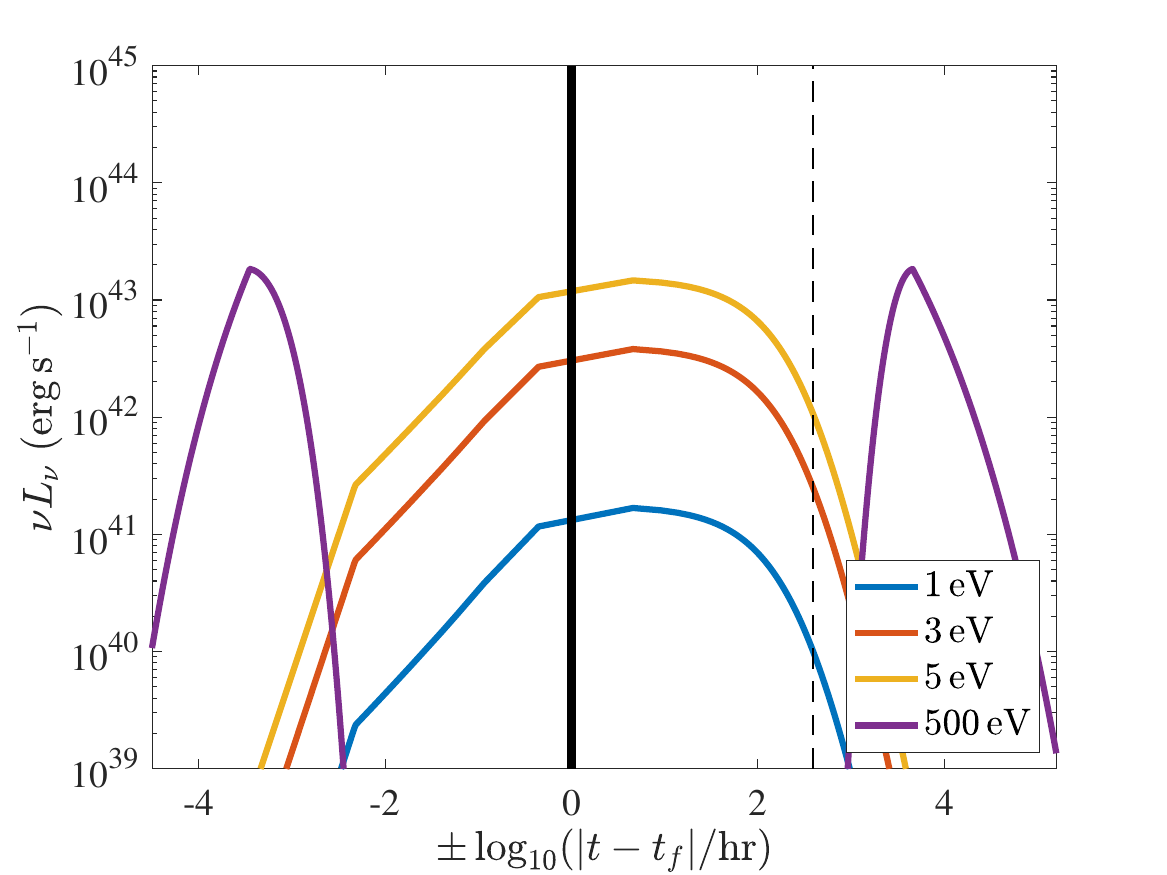}
    \caption{Band-dependent lightcurves ($\nu L_\nu$) from a sun-like star undergoing unstable mass transfer to an SMBH of mass $\MBH=10^6 \, \rm M_\odot$. Prior to the final disruption of the star, the emission is accretion dominated, peaking at X-rays, while towards the final disruption of the star at time $t_f$, super-Eddington outflow develops, temporarily producing cooler and brighter emission. Following the star's final disruption and the decay in viscously-limited accretion, the disc's emission dominates at late times. \textit{Top panel}: linear horizontal time axis, around time $t=t_f$. \textit{Bottom panel}: Same as the top panel, but with a double logarithmic horizontal axis, with negative values indicating time prior to the final disruption of the star, and vice versa. \textit{Black dashed vertical line} marks one viscous time (Eq.~\ref{eq:t_visc}) following the star's disruption.}
    \label{fig:nuLnu_vs_time}
\end{figure}

Finally, we present the band-dependent lightcurves, $\nu L_\nu(t)$ for a few characteristic frequencies in Fig.~\ref{fig:nuLnu_vs_time}. At early and late times, the emission is dominated by soft X-rays originating from the inner regions of an accretion disk, with peak luminosities comparable to $L_{\rm Edd}$. Around the time of final disruption, as super-Eddington outflows develop, the emission becomes brighter yet cooler, resulting in a substantial decay in X-ray flux. In the last weeks prior to the final disruption of the star, the emission peaks at UV/Optical/IR, which then decays and is replaced by X-rays from the inner regions of the disk as the accretion rate drops below $\dot{\mathcal{M}}_\bullet \lesssim 1$. In computing $\nu L_\nu$ we have assumed black-body emission of luminosity $L_{\rm bol}(t)$ and $\kb T_{\rm BB}(t)$. While this is a reasonable assumption for the emission produced by the expanding optically thick super-Eddington outflow, we admittedly misrepresent the multicolor emission from the resulting accretion disk, by considering only the innermost disk annuli (i.e., the temperature given by Eq.~\ref{eq:T_BB_disk}), possibly underestimating the emission at softer bands. Yet, this simplification does not result in a qualitative difference, aside from changing the slopes of the rising UV/Optical/IR fluxes in Fig.~\ref{fig:nuLnu_vs_time}. Specifically, our approximation does not capture the optical/UV plateaus seen in regular TDEs at late times, also expected to occur in circular TDEs, as the disk cools and viscously spreads \citep[e.g.,][]{Mummery+24}.

\begin{figure}
    \centering
    \includegraphics[width=0.49\textwidth]{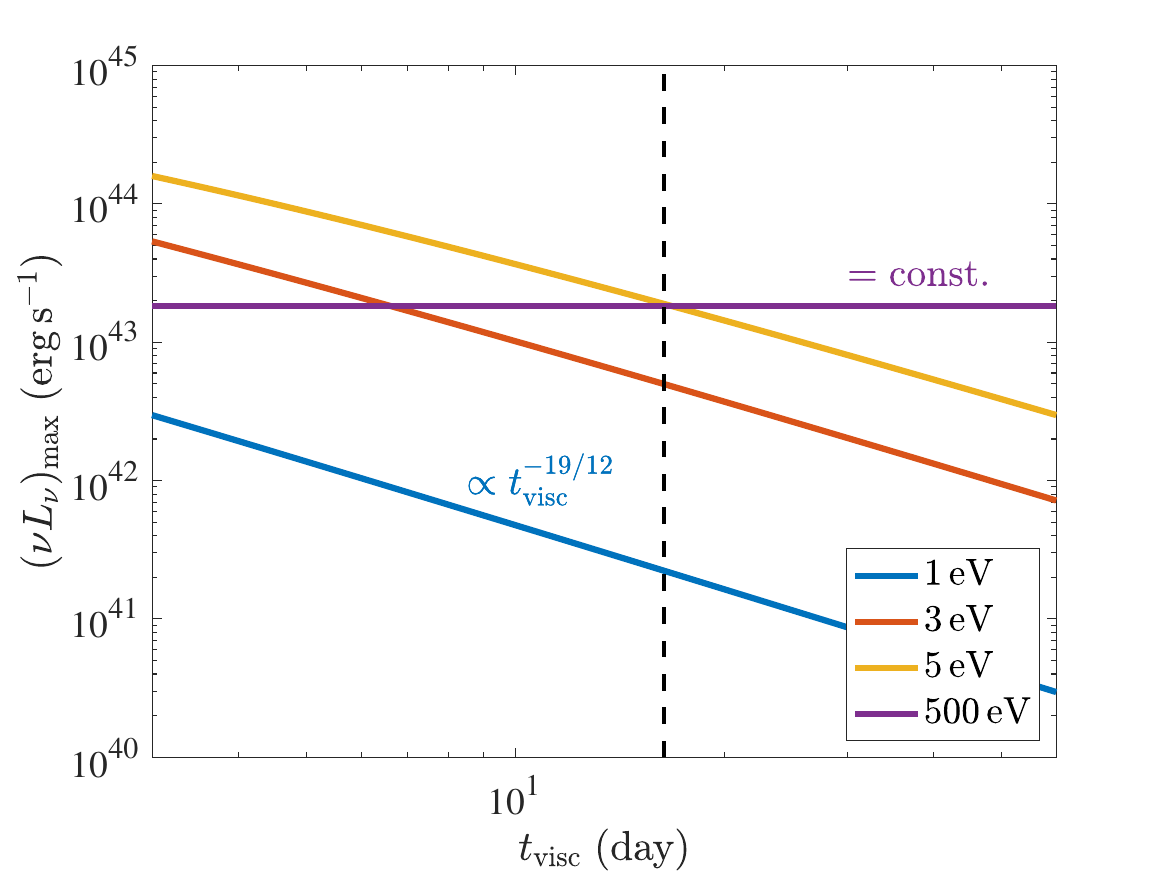}
    \includegraphics[width=0.49\textwidth]{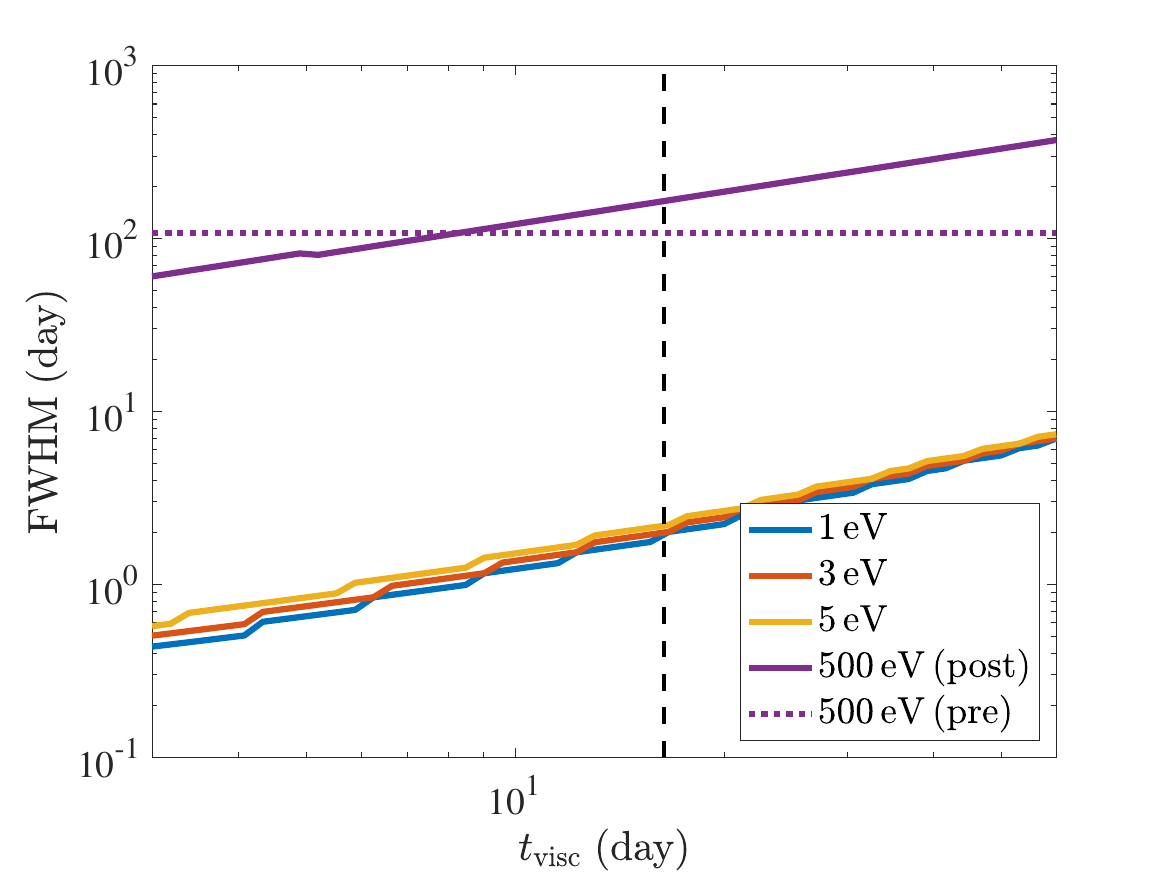}
    \caption{\textit{Top panel}: The maximum value of $\nu L_\nu$ in different observational bands, as a function of $t_{\rm visc}$, the (somewhat uncertain) viscous time at $r_{\rm MT}$ (Eq.~\ref{eq:t_visc}). \textit{Bottom panel}: The full-width at half maximum (FWHM) as a function of $t_{\rm visc}$. For $\nu=500\,\rm eV$ band, we plot the FWHM of both the pre- and post-disruption flares (see Fig.~\ref{fig:nuLnu_vs_time}) in dotted and solid purple lines, respectively. The fiducial value used throughout the paper is marked by the vertical black dashed line.}
    \label{fig:tvisc}
\end{figure}

The viscous time in the accretion flow at the radius where the star undergoes its final tidal disruption sets the overall timsecale for the accretion of the star onto the BH; it thus also sets the characteristic luminosity of the resulting transient.  Unfortunately the viscous time is not that well known a priori, and may depend on the magnetic field in the star prior to disruption.   To demonstrate how the properties of the predicted transient depend on $t_{\rm visc}$, Figure \ref{fig:tvisc} shows the peak luminosity from the IR to the X-ray, as well as the duration of the transient, as a function of the viscous time.  For somewhat longer viscous times, the optical flare predicted here is quite similar to those of observed luminous fast blue optical transients (LFBOTs), a point we return to in \S \ref{sec:lfbot}.

\begin{figure}
    \centering
    \includegraphics[width=0.5\textwidth]{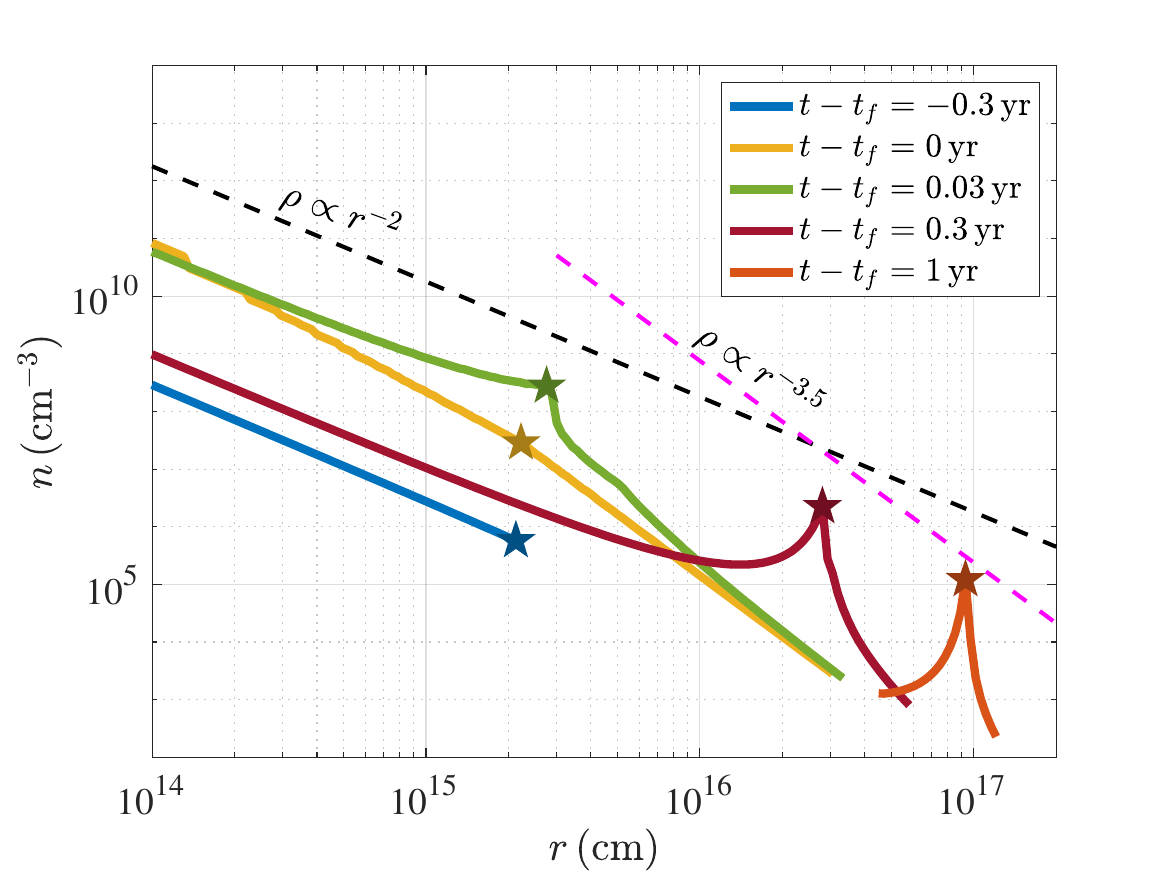}
    \caption{Circum-nuclear density produced by super-Eddington outflow from a circular-TDE, for $m_\star=1 \, \rm M_\odot$ and $\MBH = 10^6 \, \rm M_\odot$. Different colors correspond to different epochs relative to the complete disruption of the star. Stars indicate the distance at which $n r^3$ is maximized, corresponding to where most of the ejected mass is concentrated.}
    \label{fig:DensityProfiles}
\end{figure}

{Finally, we use the time-evolution of $\dot{m}_\star$ and $\dot{m}_{\rm w}$ to infer the density of the surrounding circum-nulcear medium. Under the simplified assumption of constant $\beta_{\rm w}$ and $f_{\rm w}$, the matter found at position $r$ at time $t$ is due to outflow launched from the BH vicinity at time $t_0 = t - r/(\beta_{\rm w}c)$, when the outflow was launched at rate $\dot{m}_{\rm w}(t_0)$. Assuming that strong outflows with $f_{\rm w}\approx 0.1$ are launched when $\dot{\mathcal{M}}_\bullet > 1$, we plot the density profiles at different epochs in Fig.~\ref{fig:DensityProfiles}. 
The first epoch, $t = t_f -0.3 \, \rm yr$, is very close to $t_{\rm Edd}$ (Eq.~\ref{eq:transition_to_Edd_MS}), and the mass transfer rate has not doubled yet, $1 < \dot{\mathcal{M}} \lesssim 2$ , such that the density follows a wind-like $\rho \propto r^{-2}$ profile, up to a maximal distance of $r_{\rm max} \approx v_{\rm w}(t-t_{\rm Edd}) \approx 10^{15} \, \rm cm$. The density increases as the mass transfer rate runs away, forming a steeper density profile, that tends to $\rho \propto r^{-3.5}$, with $2 + (1/(1-1/k)) = 3.5$ reflecting the divergence in $\dot{m}_{\rm w}$ towards the disruption of the star. At even later epochs, the outflow is fed by the viscously spreading accretion disk, until at times $t > t_f + \Delta t_{\rm Edd,\downarrow}$ (Eq.~\ref{eq:transition_below_Edd_MS}) no additional outflow is produced, and the ejected mass propagates as a shell traveling at velocity $v_{\rm w}$.   Figure \ref{fig:DensityProfiles} neglects outflows when $\dot{\mathcal{M}}_\bullet \lesssim 0.01$, when the flow is also radiatively inefficient, and likely to produce strong outflows.  This early-time mass-loss will not dominate the total ejecta mass but will produce a circumnuclear nuclear medium at somewhat larger radii than shown in Figure \ref{fig:DensityProfiles}.}

\section{Formation scenarios and rates}
\label{sec:Formation}
How do stars in the center of a galaxy evolve onto tightly bound, low eccentricity orbits around the central SMBH, ultimately resulting in mass transfer? Within a nuclear star cluster, the vast majority of stars are on markedly wider orbits, of $\sim \rm pc$ scale. Relaxational processes prevailing in these dense stellar environments perturb stellar orbits to low angular momentum (highly eccentric) orbits, which may then undergo circularization and inspiral through GW emission, bringing them into the tight orbits of interest.

The scattering rate of stars is typically dominated by semi-major axis comparable to the SMBH's radius of influence, or the radius enclosing a total stellar mass $\approx \MBH$
\begin{equation}
    \rh \approx \frac{G\MBH}{\sigma_{\rm h}^2} \approx 1.0 \, {\rm pc} \; M_{\bullet,6}^{0.5} \,,
\end{equation}
where $\sigma_{\rm h}$ is the cluster's velocity dispersion, and we have assumed the $\MBH$-$\sigma_{\rm h}$ relation, $\MBH \propto \sigma_{\rm h}^4$ \citep[e.g.,][]{Tremaine_2002}. Stars of semi-major axis $\approx \rh$ may undergo angular momentum relaxation, and be perturbed into the loss-cone, ultimately leading to their tidal disruption, as they approach $\rp\approx \rtidal$ on a highly eccentric, nearly parabolic orbit \citep[e.g.,][]{Rees_1988,Stone_2020}.

An alternative evolutionary scenario may occur for stars that are initially paired as binaries. Consider a binary star system of separation $a_{\rm b}$ and components of similar mass $\approx m_\star$, orbiting the SMBH at around $\rh$. The binary may be scattered onto a highly eccentric orbit, resulting in its tidal splitting through Hills' mechanism \citep[][]{Hills_1988}, ejecting one star as a hyper-velocity star, and leaving the other member on a bound orbit of pericenter distance $r_{\rm p,i} \approx a_{\rm b} (\MBH/m_{\rm \star})^{1/3}$ and semi-major axis $a_{\rm i} \approx \frac{1}{2} a_{\rm b} (\MBH/m_{\rm \star})^{2/3}$ \citep[e.g.,][]{Hills_1988,Kobayashi_2012,Cufari_2022}.

The subsequent evolution of the bound star depends primarily on the rate of orbital angular momentum diffusion through scattering processes ($\left< J/\dot{J}_{\rm rlx} \right> = \tau_{\rm rlx}^{\rm J})$, versus its orbital energy dissipation rate through GW emission ($|E_{\rm orb}/\dot{E}_{\rm GW}| = \tgw^{\rm E}$). We focus on uncorrelated, two-body scatterings as the dominant angular momentum relaxation process, yielding
\begin{equation}
    \frac{\ttb^{\rm J}}{\tgw^{\rm E}} \approx \frac{1}{\ln{\Lambda}} \pfrac{a_{\rm i}}{\rh}^{\gamma-11/2} \pfrac{R_{\rm g}}{\rh}^{5/2} \pfrac{r_{\rm p,i}}{a_{\rm i}}^{-5/2} \,,
\end{equation}
evaluated shortly after the binary undergoes tidal splitting, where $\ln{\Lambda}\approx \ln{\MBH/m_\star} \sim \mathcal{O}(10)$ is the Coulomb logarithm, and the density profile of stars around the SMBH is assumed to scale as $n_\star (r) \propto r^{-\gamma}$. Substituting $r_{\rm p,i}/a_{\rm i} \approx (m_{\rm \star}/\MBH)^{1/3}$ and considering a Bahcall-Wolf density profile, with $\gamma = 7/4$ \citep{BW_76,BW_77}, we find
\begin{equation}
    \frac{\ttb^{\rm J}}{\tgw^{\rm E}} \approx \frac{1}{\ln{\Lambda}} \pfrac{\MBH}{m_{\rm b}}^{5/6} \pfrac{R_{\rm g}}{\rh}^{5/2} \pfrac{a_{\rm i}}{\rh}^{-15/4} \,,
\end{equation}
or for fiducial values
\begin{multline}
    \frac{\ttb^{\rm J}}{\tgw^{\rm E}} \approx 
    5 \times 10^{-15} \\
    \, \pfrac{\ln \Lambda}{10}^{-1} \pfrac{m_{\star}}{\rm M_\odot}^{-5/6} M_{\bullet,6}^{2.1} \pfrac{a_{\rm i}}{\rh}^{-15/4} \,.
\end{multline}

If $\tau_{\rm GW}^{\rm E} \lesssim \tau_{\rm 2B}^{\rm J}$, the orbit evolves primarily due to GW emission, with decreasing eccentricity until the onset of mass transfer, at which point its remaining eccentricity is
\begin{equation} \label{eq:FinalEcc}
    e_{\rm f} \approx \pfrac{R_\star}{a_{\rm b}}^{19/12} \pfrac{425}{304}^{145/242} \pfrac{\amt}{2}^{19/12} \,,
\end{equation}
where we used the constant of motion obtained by \cite{Peters_1964} for quadrupole GW emission, and assumed $r_{\rm p,i} \approx \rtidal (a_{\rm b}/R_\star) \ll a_{\rm i}$, and that mass transfer ensues at a separation of $r_{\rm MT} \approx \amt \rtidal$.

Substituting $a_{\rm i} = \frac{1}{2} a_{\rm b} (\MBH/m_{\star})^{2/3}$, we find that the maximal initial binary separation for which GWs dominate over 2-body scattering for the orbit of the bound star shortly after the binary is disrupted, is approximately
\begin{multline}
    \left. \frac{a_{\rm b}}{R_\star}\right|_{\rm 2B=GW} \approx \\ 1.4 \;
    \pfrac{\ln \Lambda}{10}^{-4/15} \pfrac{\MBH}{10^6 \, \rm M_\odot}^{0.39} \pfrac{m_{\star}}{\rm M_\odot}^{4/9} \pfrac{R_\star}{1 \, \rm R_\odot}^{-1} \,.
\end{multline}
A similar formation scenario has been previously invoked as a possible origin of QPE sources, as discussed in \cite{Metzger_2021,Krolik_Linial_2022,Linial_Sari_2023,Lu_Quataert_23,Linial_Metzger_23}.

For our fiducial values, the minimal residual eccentricity at the onset of mass transfer is roughly $e_{\rm f} \approx 0.7$ (Eq.~\ref{eq:FinalEcc}), for which substantial tidal heating is expected to take place (appendix \ref{sec:AppendixTidalHeating}), leading to an earlier onset of mass transfer, with $\amt \gg 2$. Considering a somewhat more massive SMBH and a lower mass star, the remaining eccentricity at $r\approx \amt \rtidal$ is reduced to
\begin{multline}
    e_{\rm f} \gtrsim e_{\rm f,min} \approx 0.1 \\
    \pfrac{R_\star}{0.5 \, \rm R_\odot}^{19/12} \pfrac{m_\star}{0.5 \, \rm M_\odot}^{-19/27} M_{\bullet,7}^{-0.6} \pfrac{\amt}{2}^{19/12} \,,
\end{multline}
still greatly exceeding the critical eccentricity $e \approx 10^{-3}-10^{-2}$ above which tidal heating likely alters the stellar structure significantly (Appendix \ref{sec:AppendixTidalHeating}).

One possible mechanism for reducing the orbital eccentricity prior to the onset of mass transfer and/or substantial tidal heating is via hydrodynamical drag with gas present at distances from the SMBH commensurate with the stellar orbit. The source of gas could be a large scale accretion flow, typical of active galactic nuclei (AGN), as considered for example in \cite{Macleod_Lin_2020,Tagawa_23, Linial_Quataert_24}. Alternatively, drag may be induced by collisions between the star and a compact accretion disk that forms following the (parabolic) tidal disruption of another star in the same galactic nucleus. The effect of encounters between a tightly bound MS star and a TDE-like accretion disk has been recently studied in the context of QPEs \citep[e.g.,][]{Linial_Metzger_23,Franchini_23,Tagawa_23}. Depending on the efficiency of interaction between the star and a preexisting accretion flow, sufficiently low eccentricity systems may form to diminish the effects of tidal heating. However, if the star's orbit is highly inclined relative to the disk, the star may suffer substantial mass loss due to ablation occurring as the star passes through the disk \citep[e.g.,][]{Linial_Metzger_23,Linial_Quataert_24}.

The formation rates of both stellar-EMRIs and "regular" TDEs are set by the angular momentum relaxation of orbits into the loss cone, acting as the evolution bottleneck. If the fraction of stars paired in binaries at the $\rh$ is $f_{\rm b}$, and if the loss-cone is empty for both binary splitting and TDEs, their respective rates are approximately \citep[e.g.,][]{Cohn_Kulsrud_78,Merritt_2013,Broggi_2024}
\begin{multline} \label{eq:Rate_TDE}
    \left. \mathcal{R}_{\rm TDE}\right|_{\rm e.l.c} \approx \frac{N_\star(\rh)}{\ttb(\rh)} \frac{1}{\ln{\left( \rh/\rtidal \right)}} \approx \\
    \frac{\sigma_{\rm h}^3}{G\MBH} \pfrac{\ln{\Lambda}}{\ln{\left( \rh/\rtidal \right)}} \approx 10^{-4} M_{\bullet,6}^{-0.25} \,,
\end{multline}
and the stellar-EMRI rate is
\begin{multline}
    \left. \mathcal{R}_{\rm \star-EMRI} \right|_{\rm e.l.c} \approx f_{\rm b} \left. \mathcal{R}_{\rm TDE}\right|_{\rm e.l.c} (1 + \ln(a_{\rm b}/R_\star)) \approx \\
    10^{-5} \, \pfrac{f_{\rm b}}{0.1} M_{\bullet,6}^{-0.25} \,,
\end{multline}
where the mostly insignificant logarithmic factor accounts for the larger effective loss-cone associated with binary splitting. As long as the distribution of $a_{\rm b}$ among the binaries is roughly log-uniform, the distribution of eccentricities at the onset of mass transfer is expected to be fairly flat. {Tidal hardening of the binary during its orbit around the SMBH prior to disruption could, however, lead to an excess of stars on moderately eccentric orbits at RLO after Hills breakup of the binary, relative to stars with very low eccentricity \citep{Lu_Quataert_23}.} 

\section{White Dwarf and an IMBH} \label{sec:WD_Circular_TDE}
In this section we consider an analogous problem, of a white dwarf on a circular orbit, undergoing unstable mass transfer to an IMBH. A similar scenario has been explored for for example in \cite{Zalamea_2010, Xin_2024}, and here we focus on the observable properties of such system. The general arguments applied in the context of a MS star can be used here, with some important differences and subtleties which we discuss. The key qualitative differences are the much shorter dynamical time and GW inspiral timescale for a Roche-lobe filling WD. We also note that because nearly circular orbits are more readily achieved, tidal heating may play a weaker role in driving the onset of mass transfer compared with MS stars, and hence we adopt $\amt=2$ throughout this section.

The orbital period at $a=r_{\rm MT}$ is
\begin{multline}
    P_{\rm orb} \approx
    30 \, {\rm s} \; \pfrac{\amt}{2}^{3/2} \pfrac{R_\star}{0.01 \rm R_\odot}^{3/2} \pfrac{m_\star}{\rm M_\odot}^{-1/2}
    \, \approx \\
    60 \, {\rm s} \; \pfrac{\amt}{2}^{3/2} \pfrac{m_\star}{0.5 \, \rm M_\odot}^{-1} \; \rm (WD) \,. \label{eq:PtidalWD}
\end{multline}
and in the final expression we used the mass-ratio relation appropriate for a low-mass WD, $R_\star \propto m_\star^{-1/3}$, such that $P_{\rm orb} \propto m_\star^{-1}$.

{In order to avoid direct capture by the SMBH, $r_{\rm MT} \geq r_{\rm ISCO}$, implying a relatively "low-mass" SMBH and WD, as  
\begin{equation}
    \frac{r_{\rm MT}}{R_{\rm g}} \approx 1.5 \, \pfrac{m_\star}{0.5 \, \rm M_\odot}^{-2/3} \pfrac{\MBH}{10^6 \, \rm M_\odot}^{-2/3} \pfrac{\amt}{2} \; \rm (WD) \,,
\end{equation}
Thus, the condition for the WD to overfill its Roche lobe outside the innermost circular orbit (ISCO) of a Schwarzschild SMBH ($r_{\rm MT}/R_{\rm g} \gtrsim 6$) can be stated as
\begin{equation}
    m_\star \MBH \lesssim 6.2 \times 10^4 \; {\rm M_\odot^2} \; \pfrac{r_{\rm ISCO}}{6 \, R_{\rm g}}^{-3/2} \pfrac{\amt}{2}^{3/2} \, \; \rm (WD) \,.
\end{equation}
In what follows, we adopt $\MBH = 10^5 \, \rm M_\odot$ and $m_\star = 0.5 \, \rm M_\odot$ for the WD case, marginally satisfying the above criterion.}

The corresponding GW timescale is
\begin{multline} \label{eq:t_gw_wd}
    \tgw \approx \\
    200 \, {\rm d} \; M_{\bullet,5}^{-2/3} \pfrac{m_\star}{0.5 \, \rm M_\odot}^{-11/3} \pfrac{\amt}{2}^4 \; \rm (WD)
    \,,
\end{multline}
implying that for a stable mass transfer driven by GW orbital decay, the equilibrium rate greatly exceeds the Eddington accretion rate
\begin{multline} \label{eq:Mdot_eq_WD}
    \dot{\mathcal{M}}_{\rm eq,GW} \approx \frac{m_\star/\tgw}{\dot{M}_{\rm Edd}} \approx \\
    350 \; \epsilon_{-1} \, M_{\bullet,5}^{-1/3} \pfrac{m_\star}{0.5 \, \rm M_\odot}^{14/3} \pfrac{\amt}{2}^{-4} \; \rm (WD) \,,
\end{multline}
However, for the reasons discussed in the context of MS stars, the mass transfer from a WD will be highly non-conservative (in terms of the orbital angular momentum), and we thus anticipate  unstable, runaway evolution. 

While the MS scenario is characterized by a long lived secular phase, with $\dot{\mathcal{M}}_\star \ll 1$, here high mass loss rates are obtained shortly after the onset of mass transfer. Mass loss at the Eddington rate is first obtained during the GW driven phase, at time
\begin{multline}
    t_{\rm Edd} \approx \tgw \pfrac{\dot{M}_{\rm Edd} \tdyn}{m_\star}^{1/k} \approx \\
    3.9 \, {\rm hr} \, \epsilon_{-1}^{-1/3} M_{\bullet,5}^{-1/3} \pfrac{m_\star}{0.5 \, \rm M_\odot}^{-13/3} \pfrac{\amt}{2}^4  \; \rm (WD) \,,
\end{multline}
after the onset of mass transfer (and compare with \citealt{Zalamea_2010}). $\dot{\mathcal{M}}_\star \approx 0.01$ is achieved at a somewhat earlier time $t_{\rm Edd} (0.01)^{1/k} \approx 0.8 \rm \, hr$. However, it is unlikely that a quasi-steady flow can develop sufficiently fast, during merely $\mathcal{O}(10)$ orbital periods \citep[e.g.,][]{Chashkina_24}, considering the viscous time at $r_{\rm MT}$
\begin{multline} \label{eq:t_visc_wd}
     t_{\rm visc} \approx \frac{P_{\rm orb}}{2\pi} \pfrac{h}{r_{\rm MT}}^{-2} \alpha^{-1} \approx \\
    0.3 \, {\rm hr} \, \pfrac{h/r_{\rm MT}}{0.3}^{-2} \alpha_{-1}^{-1} \pfrac{m_\star}{\rm 0.5 \, M_\odot}^{-1} \pfrac{\amt}{2}^{3/2} \, {\rm (WD)} \,.
\end{multline}

The accretion onto the IMBH from a WD EMRI commences only fairly close to $\dot{\mathcal{M}}_\star \lesssim 1$, when the evolution time becomes comparable to (or longer than) $t_{\rm visc}$. As mass loss and accretion continue to increase to super-Eddington rates, the emitted luminosity is approximated by the spherical outflow picture we discussed in \S \ref{sec:EM_from_wind}. An important caveat in the WD case is that because the WD is disrupted so close to the BH, the fraction of mass lost to an outflow and the speed of the outflow are more uncertain than for a main sequence star disrupted well outside the ISCO.

The runaway evolution of the WD dominates over GW inspiral at time (compare with Eq.~\ref{eq:tau_runaway_MS})
\begin{equation}
    t_{\rm runaway} \approx 1 \, {\rm d} \, M_{\bullet,5}^{-4/9} \pfrac{m_\star}{0.5 \, \rm M_\odot}^{-25/9} \pfrac{\amt}{2}^{8/3} \; (\rm WD) \,,
\end{equation}
after mass transfer first began. Given the short viscous timescale, the peak feeding Eddington ratio is approximately
\begin{multline}
    \dot{\mathcal{M}}_{\rm visc,0} \approx
    6\times 10^6 \\ 
    \epsilon_{-1} \pfrac{\amt}{2}^{-3/2} \alpha_{-1} \pfrac{h/r}{0.3}^2 M_{\bullet,5}^{-1} \pfrac{m_\star}{0.5 \, \rm M_\odot}^{2} \; {\rm (WD)} \,,
\end{multline}
decaying below $\dot{M}_{\rm Edd}$ at time
\begin{multline}
    \Delta t_{\rm Edd,\downarrow} \approx 10 \, {\rm d} \,
    \pfrac{h/r}{0.3}^{-5/4} \\
    \alpha_{-1}^{-5/8} \pfrac{\amt}{2}^{15/16} \pfrac{m_\star}{0.5 \, \rm M_\odot}^{-1/4} M_{\bullet,5}^{-3/8} \, \rm (WD) \,
\end{multline}
after the disruption of the star. The peak luminosity associated with super-Eddington outflow produced at this phase is approximately
\begin{multline}
    L_{\rm bol}^{\rm max} \approx 2\times 10^{45} \, {\rm erg s^{-1}} f_{\rm w}^{1/3} \alpha_{-1}^{1/3} \beta_{\rm w}^{2/3} \\
    \pfrac{\amt}{2}^{-1/2} \pfrac{h/r}{0.3}^{2/3} M_{\bullet,5}^{2/3} \pfrac{m_\star}{0.5 \, \rm M_\odot}^{2/3} \; \rm (WD) \, ,
\end{multline}
comparable to the peak luminosity obtained in the MS case, yet a much smaller total radiated energy at around peak
\begin{multline}
    L_{\rm bol}^{\rm max} \times t_{\rm visc} \approx 5\times 10^{48} \, {\rm erg} \;
    f_{\rm w}^{1/3} \beta_{\rm w}^{2/3} \, M_{\bullet,6}^{2/3} \\
    \pfrac{h/r_{\rm MT}}{0.3}^{-4/3} \pfrac{\alpha}{0.1}^{-2/3} \pfrac{m_\star}{\rm M_\odot}^{-1/3} \pfrac{\amt}{2}^{1/6} \, {\rm (WD)} \,.
\end{multline}
and a corresponding blackbody temperature
\begin{multline}
    \kb T_{\rm BB} (L_{\rm bol}^{\rm max}) \approx 0.15 \, {\rm eV} \; f_{\rm w}^{-5/12} \beta_{\rm w}^{-1/12} \, M_{\bullet,5}^{1/6} \\
    \pfrac{\amt}{2}^{15/24} \pfrac{\alpha}{0.1}^{-5/12} \pfrac{h/r}{0.3}^{-5/6} \pfrac{m_\star}{0.5 \, \rm M_\odot}^{-5/6} \; \rm (WD) \,.
\end{multline}

\begin{figure*}
    \centering
    \includegraphics[width=0.49\textwidth]{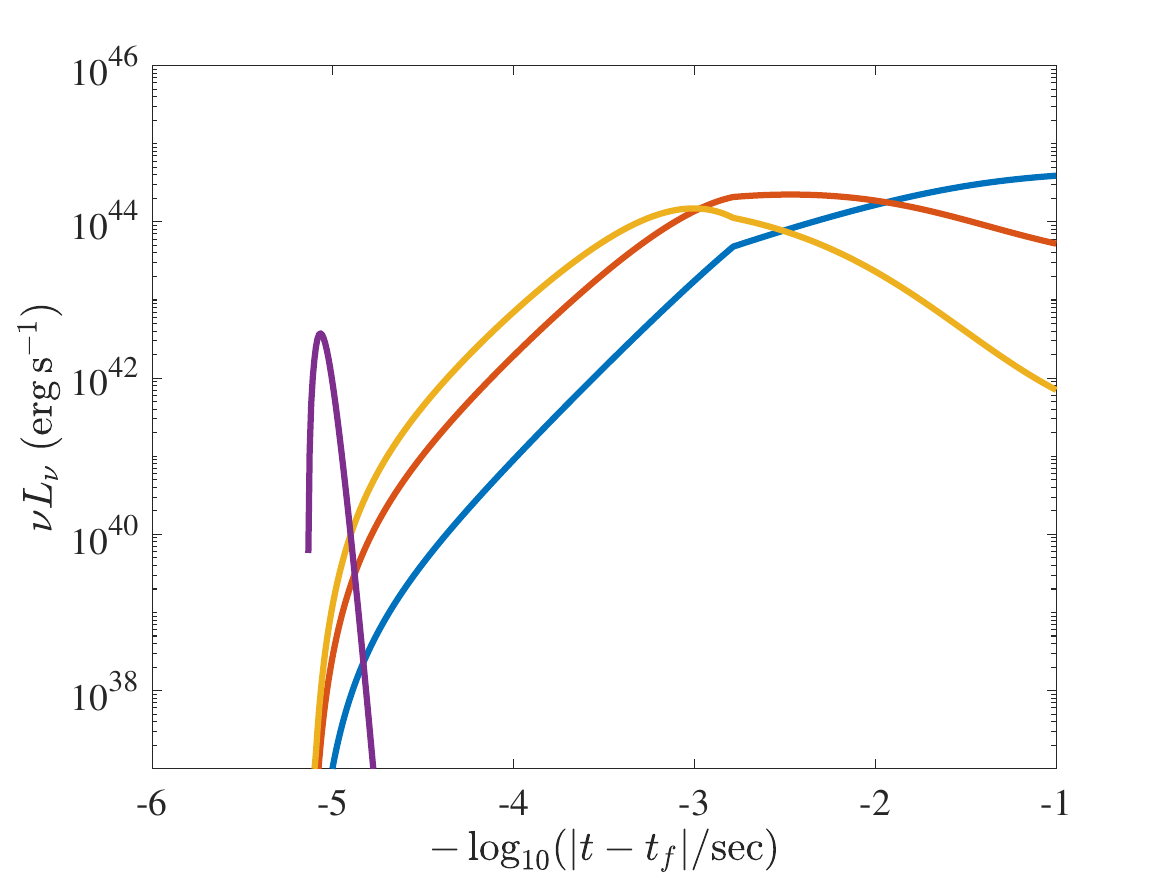}
    \includegraphics[width=0.49\textwidth]{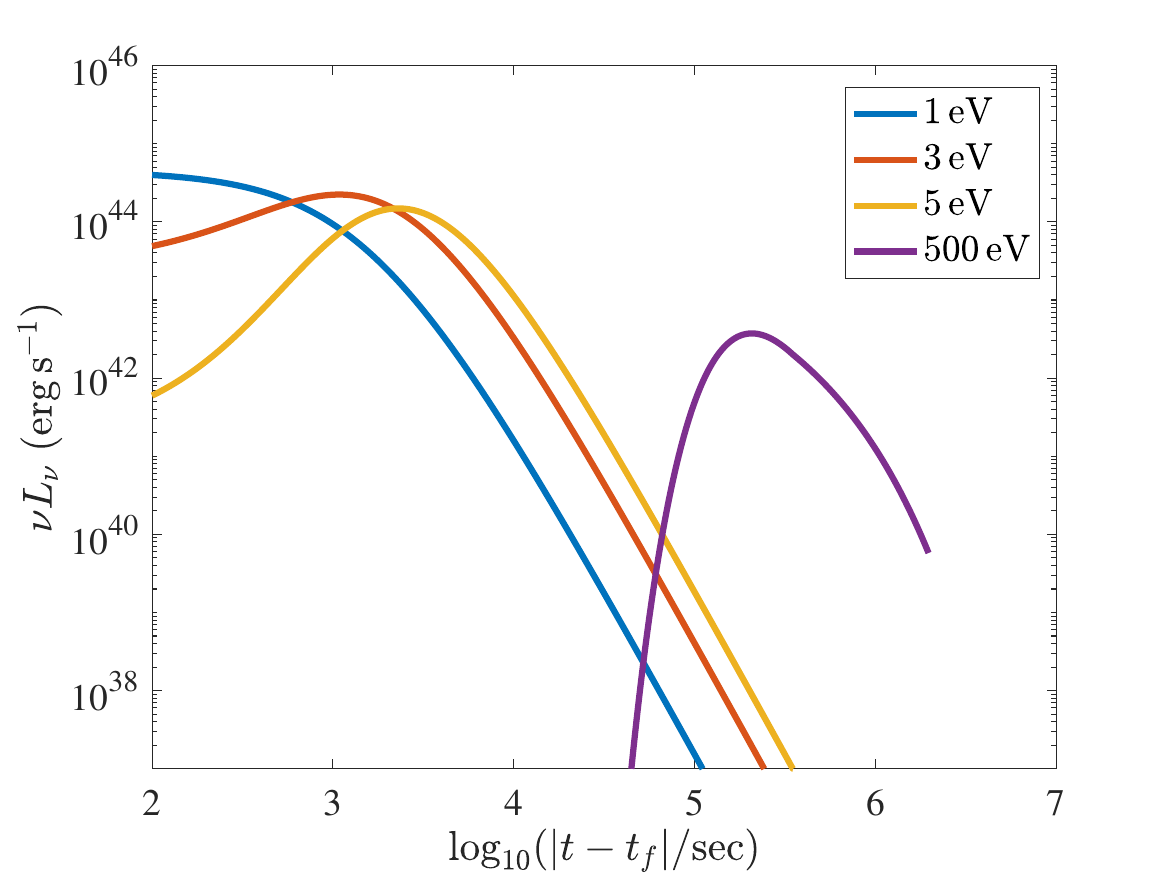}
    \caption{Band dependent lightcurve, $\nu L_\nu$ for a $m_\star=0.5 \, \rm M_\odot$ white dwarf and an $\MBH=10^5 \, \rm M_\odot$ IMBH.}\label{fig:nuLnu_vs_time_WD}
\end{figure*}

We demonstrate the band-dependent emission from a WD undergoing unstable mass transfer to an IMBH in Fig.~\ref{fig:nuLnu_vs_time_WD}. Again, we see a phase of soft X-ray emission before and after the disruption of the WD, with a UV/optical/IR emission during a short phase of super-Eddington outflow resulting around the disruption of the WD.

For a WD binary, the minimal eccentricity at the onset of mass transfer following Hills' mechanism is much smaller, at approximately
\begin{equation}
    e_{\rm f,min} \lesssim 10^{-4} \pfrac{\MBH}{10^5 \, \rm M_\odot}^{-0.36} \,,
\end{equation}
precluding substantial tidal heating prior to the onset of RLO (Appendix \ref{sec:AppendixTidalHeating}).

The band-dependent lightcurve of our fiducial WD-IMBH circular TDE is shown in Fig.~\ref{fig:nuLnu_vs_time_WD}, pre- and post disruption (left and right subpanels, respectively). 

\subsection{GW emission}
The inspiral of a WD towards an SMBH produces GWs of frequency
\begin{multline}
    f_{\rm GW} \approx \frac{2}{P_{\rm orb}} = \frac{1}{\pi \amt^{3/2}} \tdyn^{-1} \approx \\
    35 \, {\rm \, mHz} \, \pfrac{m_\star}{0.5 \, \rm M_\odot} \pfrac{\amt}{2}^{-3/2} \; {\rm (WD) } \,,
\end{multline}
where we assumed $a \approx r_{\rm MT}$. This frequency is
well within the peak-sensitivity range of LISA, a future space-based GW antenna \citep[e.g.,][]{Amaro_Seoane_2018}. The (dimensionless) characteristic strain of such a source is enhanced by the square-root of the number of GW cycles completed over time $T$, $\sqrt{f_{\rm GW} T}$, where $T$ is the shorter of the observation baseline (limited by the mission's duration) and the source's coherence time, i.e., $\tgw$. For the fiducial parameters we considered, the source evolves over at most $\tgw \approx 200 \, \rm d$ (Eq.~\ref{eq:t_gw_wd}), or even shorter durations once unstable mass transfer ensues, such that $\sqrt{f_{\rm GW} T} \lesssim 780$. For a source at distance $d$, the characteristic strain is \citep[e.g.,][]{Robson_2018}
\begin{multline}
    h_c \approx \frac{8}{\sqrt{5}} \frac{R_{\rm g}}{r_{\rm MT}} \frac{Gm_\star/c^2}{d} \sqrt{f_{\rm GW} \tgw} \approx \\
    10^{-17} \; \pfrac{d}{\rm Mpc}^{-1} \pfrac{\MBH}{10^5 \, \rm M_\odot}^{1/3} \, \pfrac{m_\star}{0.5 \, \rm M_\odot}^{-1/3} \pfrac{\amt}{2}^{1/4} \,,
\end{multline}
corresponding to an SNR of order $\mathcal{O}(10^3)\times (d/\rm Mpc)^{-1}$, where the above expression is valid as long as $\tgw \lesssim 5 \, \rm yr$ \citep{Amaro_Seoane_2018,Robson_2018}. For a limiting ${\rm SNR} \gtrsim C$, approximately $N_{\rm gal}({\rm SNR} > C) \approx (4\pi/3) n_{\rm gal}(\MBH) d_{\rm SNR=C}^3 \approx 5\times 10^4 \, (C/10)^{-3}$ galaxies hosting an SMBH of mass $\MBH \approx 10^5 \, \rm M_\odot$ are expected to be contained within a detection horizon of roughly $d_{\rm SNR=C} \approx 100 \, \rm Mpc \; (C/10)^{-1}$. 

Assuming steady state, the production of newly formed WD-EMRIs is compensated by their disruption. Thus, the number of systems expected to display both a detectable GW signal and an EM counterpart is roughly
\begin{multline}
    N_{\rm src}({\rm SNR>C}) \approx
    N_{\rm gal} \mathcal{R}_{\rm WD-EMRI} T_{\rm LISA} \approx \\
    \mathcal{O}(1) \pfrac{\mathcal{R_{\rm WD-EMRI}}}{10^{-5} \, \rm yr^{-1} \, gal^{-1}} \pfrac{T_{\rm LISA}}{5 \, \rm yr} \pfrac{C}{10}^{-1}\,,
\end{multline}
where $\mathcal{R}_{\rm WD-EMRI}$ is the formation rate of WD-EMRIs, which we normalized by according to the relaxation rate at the radius of influence, $\rh$ of an SMBH with $\MBH=10^5 \, \rm M_\odot$. The number of sources grows with the mission lifetime, $T_{\rm LISA}$. We therefore note that such WD-EMRIs could be detected simultaneously as multi-messenger sources both in mHz GWs, as well as a multiband electromagnetic flare.

\section{Summary and Discussion} \label{sec:Discussion}

We studied a novel channel of stellar destruction in galactic nuclei, involving the unstable Roche lobe overflow of a main-sequence star orbiting an SMBH on a nearly circular orbit. Unlike the stable mass-transfer considered in \cite{Dai_Blandford_2013} and \cite{Linial_Sari_2017}, here we have argued for an unstable evolution, expected to occur for relatively low-mass stars with convective envelopes (see also \citealt{Linial_Sari_2023,Lu_Quataert_23}). While stable, GW-driven mass transfer results in modest accretion rates ($\lesssim 10^{-5} \, \dot{M}_{\rm Edd}$, Eq.~\ref{eq:Mdot_eq_MS}), unstable, runaway RLO produces an ever increasing mass transfer rate up to the star's final disruption, limited only by the viscous evolution of the resulting accretion disk, with $\dot{M}_{\rm max} \gtrsim 10^{3} \, \dot{M}_{\rm Edd}$. The transition from sub- to super-Eddington rates, and the subsequent decay in accretion as the remaining stellar debris accretes onto the central BH, produces a complex electromagnetic transient. We find a decade-long, soft X-ray precursor originating from a thin accretion disk, with luminosity approaching $L_{\rm Edd}$.  During this phase the lightcurve may be variable on timescales set by the orbital period $\sim$ hrs due to the finite eccentricity of the star, circularization shocks, star-disk collisions, and other processes. The variability amplitude is likely to be small given that the viscous time is long compared to the orbital period (for further discussion see \citealt{Metzger_2021,Krolik_Linial_2022,Linial_Sari_2023,Lu_Quataert_23}). As the accretion rate continues to increase during the final months-year prior to the star's full disruption, an extended super-Eddington, optically thick outflow develops, resulting in an extended photosphere, producing a bright UV/optical flare, lasting days-weeks. Following the star's full disruption, the feeding rate subsides, and the SMBH is fed by the viscously spreading stellar remnants, producing soft X-rays, decaying over the course of a year.

We also considered the analogous RLO of a WD onto an IMBH, which undergoes similar qualitative evolution, albeit with considerably shorter timescales. We find luminosities of order $10^{42-44} \, \rm erg \, s^{-1}$ are obtained for duration ranging from hours-day. These sources produce mHz GWs, expected to be detectable with future space-based GW detectors up to distances of order $\approx 100 \, \rm Mpc$ \citep{Zalamea_2010}. Given their optical luminosities, such WD-IMBH systems could be promising candidates for simultaneous multi-messenger detection in the era of LISA and wide-field, high cadence transient surveys.

In the following subsections we discuss a few additional implications and broader connections between our work and other observations.

\subsection{Circular vs. Parabolic Tidal Disruption Events}

The ``circular TDEs'' considered here bear several key similarities to the standard tidal disruption of a star by an SMBH, on a highly eccentric/parabolic orbit. From a theoretical standpoint, both ``regular/parabolic'' and ``circular'' TDEs involve the disruption of a star through the action of the SMBH's tidal field, occurring over just a few stellar dynamical times, $\tdyn$. The resulting emission is powered by the conversion of the star's rest-mass energy into radiation, with some (time-varying) efficiency. Both phenomena involve an evolution from super- to sub-Eddington mass accretion rates onto the SMBH, producing complex time evolving lightcurves and spectra.

Circular and parabolic TDEs share several observational similarities as well. Much like regular TDEs, circular-TDEs appear brightest in optical/UV bands, with luminosities that are typically commensurate with (or somewhat brighter than) parabolic TDEs (within our model's uncertainties), of order $L_{\rm Edd}$. This bright optical/UV flare persists for a typical timescales of order weeks, somewhat shorter than the fallback time of the most bound stellar debris that follows the disruption of a star on a parabolic orbit ($\sim$month). We find that the circular-TDE peak flare is generally blue (possibly peaking in the EUV, Eq.~\ref{eq:kT_peak_MS}), similarly to what is seen in regular TDEs. The late time behavior of both TDE flavors is also qualitatively similar - as the optical/UV emission subsides, the emission is then dominated by X-rays powered by accretion of the remaining stellar debris onto the SMBH, decaying with time as the disk viscously evolves. Optically selected circular TDEs may thus be misclassified as regular TDEs, or other forms of AGN activity. 

One important theoretical difference between these two classes concerns the temporal evolution of accretion onto the SMBH. In a parabolic TDE, the feeding rate is initially limited by the fallback of marginally bound stellar debris, with the characteristic $t^{-5/3}$ power-law decay \citep{Rees_1988,Phinney_1989}. In circular TDEs, the victim star is initially on a tightly bound orbit at $r_{\rm MT} = \amt \rtidal$, with black hole feeding driven by mass transfer from the Roche-lobe overfilling star and viscous angular momentum transport. We note that the corresponding peak accretion rate in a circular-TDE is coincidentally not too different from the peak fallback rate occurring in regular TDEs, with their ratio given by
\begin{multline}
    \frac{\dot{m}_{\rm TDE}^{\rm max}}{\dot{m}_{\rm C-TDE}^{\rm max}} \approx \frac{t_{\rm visc}}{3 \, t_{\rm fb}} \approx \\
    \frac{0.1}{\alpha_{-1}} \; \pfrac{h/r_{\rm MT}}{0.3}^{-2}  \pfrac{\amt}{4}^{3/2} \pfrac{m_\star}{\rm M_\odot}^{1/2} M_{\bullet,6}^{-1/2} \; \rm (MS) \\
    \frac{0.1}{\alpha_{-1}} \; \pfrac{h/r_{\rm MT}}{0.3}^{-2} \pfrac{\amt}{2}^{3/2} \pfrac{m_\star}{0.5 \, \rm M_\odot}^{1/2} M_{\bullet,5}^{-1/2} \; \rm (WD) \,,
\end{multline}
where $t_{\rm fb} \approx \pi \tdyn (\MBH/m_\star)^{1/2}$ is the fallback time of the most bound debris following the disruption of a star on a parabolic orbit, and $\dot{m}_{\rm TDE}^{\rm max} \approx m_\star/(3\,t_{\rm fb})$ \citep[e.g.,][]{Guillochon_2013}.

Another key observational difference between these two flavors of stellar destruction, is that circular-TDEs are characterized by a long-lasting precursor prior to the final disruption of the star, initiated at the onset of mass transfer and corresponding accretion onto the SMBH. For MS stars, substantial disk emission at a fraction of the Eddington luminosity begins years-decade prior to the final disruption, and super-Eddington outflows commence months-year before the star is fully destroyed (Eq.~\ref{eq:transition_to_1pEdd_MS},~\ref{eq:transition_to_Edd_MS}). Thus, early X-ray emission at a fraction of $L_{\rm Edd}$, preceding the optical/UV flare may be a unique signature of circular TDEs. This also highlights the importance of not selecting against previous nuclear X-ray evidence for an AGN in `TDE' searches; such an observational selection could exclude circular TDEs.   We expect circular TDEs to be quite distinct relative to normal AGN in the year-decade leading up to stellar disruption, with bright and soft X-ray emission but comparatively little optical emission, no broad line region, and at most a faint narrow line region confined to close to the SMBH.

\subsection{Circular TDEs vs. QPEs}
Stars on tight, mildly eccentric orbits around SMBHs have been recently invoked as a potential channel for producing Quasi-Periodic Eruptions (QPEs), detected in soft X-rays \citep{Krolik_Linial_2022,Linial_Sari_2023,Lu_Quataert_23,Linial_Metzger_23,Franchini_23, Tagawa_23}. Specifically, \cite{Linial_Metzger_23} have demonstrated that a star on a few hours-day orbital period, colliding twice per orbit with an inclined accretion disk, produces bright and hot flares that may outshine the disk emission in the $0.2-2\, \rm keV$ band, consistent with the observed properties of QPE sources. 

Given the overall similarity between typical QPE periodicites and the orbital period at $\rtidal$ (of a few hours, Eq.~\ref{eq:PtidalMS}), \cite{Linial_Metzger_23} have further suggested that the origin of the impacted disk in their model is the tidal disruption of a second star, producing a compact accretion. Comparison of TDE rates and GW inspiral timescales establishes that a coincidence between such stellar-EMRIs and TDEs is likely unavoidable - stars orbiting the SMBH just outside $r_{\rm MT}$ migrate on a timescale $\tgw \approx 10^{6-7}$ yr (Eq.~\ref{eq:t_gw_ms}), much longer than the typical interval between consecutive TDEs, $\mathcal{R}_{\rm TDE}^{-1} \approx 10^4 \, \rm yr$ (Eq.~\ref{eq:Rate_TDE}).

Within this picture, circular TDEs and QPEs represent different evolutionary endpoints of stellar-EMRIs produced in galactic nuclei. If QPEs indeed occur when slowly migrating stellar-EMRIs are intercepted by the accretion disks that follow parabolic TDEs, the existence and rates of circular TDEs will depend on the survivability of a star following a phase of disk interaction. As discussed in \cite{Linial_Metzger_23,Linial_Quataert_24}, and in Yao \& Quataert (in prep.) star-disk collisions play an important in the ablation of the impacting star's outer layers. We discuss possible scenarios based on the ablation efficiency and the stellar-EMRI survivability.

One possibility is that a single encounter between a stellar EMRI (at a few $r_{\rm MT}$) and a TDE disk suffices to entirely destroy the star. \cite{Linial_Metzger_23,Linial_Metzger_24} have estimated that the collisions required to produce the observed QPE flares also result in substantial stellar ablation, occurring on a timescale that is shorter than the orbital migration time (either due to disk-induced drag or GWs). Given the above argument, $\tgw \mathcal{R}_{\rm TDE} \gg 1$, essentially all stellar-EMRIs coincide with a TDE, resulting in a phase of QPE activity, with circular TDEs being exceedingly rare. It is, however, possible that a small stellar core does survive the disk interaction, continues to inspiral until it fills is Roche lobe and produces a circular-TDE, which will follow a phase of QPE activity.

Another possible outcome is that as the stellar-EMRI interacts with the TDE disk (produced by second star's, parabolic TDE), it begins to simultaneously shed mass towards the SMBH, triggering a circular-TDE. This could occur, for example, if collisional energy deposited in the stellar envelope causes the star to to inflate up until it overfills its Roche lobe.

Stellar-EMRIs may survive the multiple TDEs they encounter as they migrate from say $2r_{\rm MT}$ to $r_{\rm MT}$ if the ablation efficiency is sufficiently low and/or the viscous evolution of the TDE disk is relatively rapid, such that the ablation time is longer than the migration time due disk-induced drag or the timescale for the TDE disk to disappear due to viscous spreading. Stellar-EMRIs then experience multiple TDEs mostly unscathed, until they eventually approach $r_{\rm MT}$ resulting in a circular TDE.

\subsection{Connection to Jetted-TDEs}
\label{sec:jet}
A small fraction of TDEs appear to produce strong relativistic jets as indicated by their radio emission, highly variable non-thermal X-ray (and even gamma-ray) emission, and afterglows \citep{Bloom2011,Andreoni2022}.  The beaming corrected fraction of jetted TDEs is $\sim 1\%$, with large uncertainties due to small-number statistics and uncertain beaming corrections.    It is unclear what distinguishes this small population of jetted TDEs from the bulk of the TDE population \citep[e.g.,][]{Teboul_Metzger_23}.   

Theoretically, an important challenge in understanding jetted TDEs is that the magnetic flux required to power a jet similar to that observed in J1644 is $\sim 10^{30} {\rm \, G \, cm^2}$ \citep[e.g.,][]{Tchekhovskoy_14} which corresponds to a coherent stellar magnetic field of $\sim 10^8$ G, much larger than the magnetic fields of even magnetic stars.   Although an accretion disk dynamo origin of this magnetic flux cannot be ruled out \citep[e.g.,][]{Liska2020}, it is also possible that the small rate of jetted TDEs reflects a second TDE channel, such as the circular TDEs considered here.   Circular TDEs are attractive candidates for producing relativistic jets because of the unusual stellar evolution leading to breakup.   The near-breakup rotation and strong tidal deformation of the progenitor star naively appear conducive to generating strong magnetic fields.   The magnetic flux could in principle approach the maximum allowed by the self-gravity of the star, $\Phi_B \sim \sqrt{6 \pi^2 G m_\star^2} \sim 4 \times 10^{30} \, (m_\star/M_{\rm \odot}) \, {\rm G \, cm^2}$.   This is sufficient to power a jet similar to that observed in relativistic TDEs.

The bright on-axis jet emission would make it challenging to detect the thermal signatures of circular TDEs highlighted in the bulk of this paper.   However, if circular TDEs indeed account for jetted TDEs, then the thermal signatures predicted here should be detected in $\sim 1 \%$ of TDEs, given beaming corrected jetted-TDE rates; e.g., $\sim 1\%$ of optically selected TDEs should have bright pre-TDE X-ray emission.

\subsection{Radio Emission and the Circum-nuclear medium}
\label{sec:CNM}
The long phase of increasing $\dot m_\star$ {\em prior} to the final disruption of the star is likely to lead to a dense circum-nuclear medium (CNM) into which later outflows will propagate, as summarized in Fig.~\ref{fig:DensityProfiles}.   This is a natural environment for producing strong radio emission.   Although the exact radio fluxes will depend on whether or not a relativistic jet is present and the relative velocity between late and early-time outflows, we can more confidently estimate the CNM properties expected since these are primarily set by the dynamical increase of $\dot m_\star$ prior to disruption.   To order of magnitude, we expect outflows to be present for at least the last $\sim$ year prior to disruption, when the accretion rate is super-Eddington (eq. \ref{eq:transition_to_Edd_MS}). As demonstrated in Fig.~\ref{fig:DensityProfiles}, near the final disruption of the star, when Eddington rates of up to $\dot{\mathcal{M}}_\bullet \sim 10^3$ are achieved, a wind-like CNM density profile containing $\dot{M}_{\rm Edd} (t_f-t_{\rm Edd}) \approx 0.01 f_w M_\odot$, and extending out to distances of up to $v_{\rm w} (t-t_{f}) \gtrsim \beta_{\rm w} \, 10^{17} \, \rm cm$, is present.  These CNM conditions are in fact intriguingly similar to those inferred for the relativistic TDE AT2022 cmc (see, e.g., Fig. 2 of \citealt{Matsumoto23}).  In \textit{Swift} J1644+57, however, the inferred CNM density extends to $10^{18} \, \rm cm$ \citep{Eftekhari2018}, requiring multiple years of outflow prior to the complete disruption of the star in our model.  This is most likely to be produced by the sub-Eddington radiatively inefficient phase of accretion prior to full disruption that we have not included in Figure \ref{fig:DensityProfiles}.

\subsection{Circular TDEs vs. Luminous FBOTs}
\label{sec:lfbot}

The model developed here for circular TDEs makes predictions that bear some resemblance to the class of luminous fast blue optical transients (LFBOTs), the prototype of which is AT2018cow \citep{Margutti2019,Perley2019,Ho2019}.   The optical flares produced by circular TDEs around SMBHs can have luminosities and timescales similar to those of LFBOTS (Fig. \ref{fig:tvisc}), and the self-generated CNM and possible jet that we predict (\S \ref{sec:jet} \& \ref{sec:CNM}) would produce nonthermal emission across the electromagnetic spectrum, qualitatively similar to what is seen in LFBOTs.   AT2018cow even has late-time UV and X-ray emission years after the initial transient detection that is consistent with continued accretion onto a central compact object \citep{Inkenhaag2023,Chen2023,Migliori2024}.   While the off-nuclear location of AT2018cow rules out its association with an SMBH, many of the models for its unusual properties invoke a stellar merger with a BH \citep{Metzger2022} or a TDE by a solar mass or intermediate mass BH \citep{Perley2019,Kremer2021}.  The qualitative similarity between our predictions and observations of LFBOTs may then indicate that there is a continuum of events with some overlap in observational properties, from the circular TDEs around SMBHs predicted here to the disruption of stars by a much lower mass BH.  The common physics responsible for this similarity would be the presence of super-Eddington accretion onto a BH in these diverse environments.  Within this broad class of events, BHs fed by stars on bound orbits (as in our circular TDE model or stellar merger models for LFBOTs) are the ones likely to produce their own CNM via mass transfer prior to full stellar disruption. It is possible that some (off-nuclear) LFBOT-like transients could be produced by the tidal disruption of a star on a roughly circular orbit around an IMBH.

\begin{acknowledgments}
We thank Brian Metzger for useful conversations and thank all of the KITP TDE24 program participants for a stimulating meeting. IL acknowledges support from a Rothschild Fellowship and The Gruber Foundation. This work was supported in part by a Simons Investigator grant from the Simons Foundation (EQ).   This research benefited from interactions at workshops funded by the Gordon and Betty Moore Foundation through grant GBMF5076, and through interactions at the Kavli Institute for Theoretical Physics, supported by NSF PHY-2309135.
\end{acknowledgments}

\appendix

\section{Tidal Heating Prior to Mass Transfer} \label{sec:AppendixTidalHeating}

In this Appendix we collect some results on the tidal heating of stars close to supermassive black holes. We begin by considering stars on a circular orbit and then quantify the eccentricity above which tidal heating can be important enough to significantly change the structure of the star at radii larger than the nominal onset of mass transfer at $\simeq 2 \rtidal$.  We show that for circular orbits tidal heating is unlikely to significantly change the structure of the star prior to RLO.  However, this is only true for $e \rightarrow 0$.   For finite eccentricity, tidal heating strongly modifies the stellar structure, leading the star to expand and initiate RLO when $r_p \sim 4-5 \rtidal$; this corresponds to $\alpha_{\rm MT} \sim 4-5$ in the main text.

\subsection{Circular Orbits}

Gravitational wave inspiral of a star on a circular orbit leads to a finite difference between the stellar spin frequency and orbital frequency even in the presence of tidal dissipation driving the star towards synchronous rotation.   This in turn leads to finite tidal heating of the star.  In this sub-section we evaluate the magnitude of this heating and its possible impact on the pre mass-transfer stellar structure for both a white dwarf and a sun-like star.

\subsubsection{White Dwarfs}
\label{sec:WDe0}
\citet{Fuller2012} and \citet{Burkart2013} calculate the tidal dissipation in a white dwarf inspiraling on a circular orbit due to the excitation of internal gravity waves.   \citet{Fuller2014} show that the influence of rotation on the excitation and damping of the waves is small for the approximately synchronously rotating case of interest here.  Absent tidal heating, mass transfer is initiated when the semi-major axis is $a \simeq r_{\rm MT} \simeq 2 (\alpha_{\rm MT}/2) \rtidal$  or equivalently when the orbital frequency is comparable to the stellar dynamical frequency.  In what follows in this section we consider a $0.5 M_\odot$ WD and do not keep the dependence on the WD mass and radius because the uncertainties in the tidal physics are much larger than the dependence on the WD properties.    

Equation 91 of \citet{Fuller2012} implies a tidal heating rate of 
\begin{equation}
\dot E_{\rm heat} \sim
3 \times 10^{39} \, {\rm erg \, s^{-1}} \, f^{-1/5} \, \left(\frac{\MBH}{10^5 \, M_\odot}\right)^{4/5} \pfrac{a}{r_{\rm MT}}^{-27/5} \pfrac{\amt}{2}^{-27/5}
\label{eq:EdotWD}
\end{equation}
The dimensionless factor $f$ depends on the WD thermal and compositional structure (which influence the buoyancy frequency and thus the excitation and propagation of internal gravity waves).  $f \sim 10-10^3$ for longer orbital periods $\gg \tau_{\rm dyn}$ but must approach $\lesssim 1$ as $P_{\rm orb} \sim \tau_{\rm dyn}$.    In more detail, the tidal heating rate in \citet{Fuller2012} and \citet{Burkart2013}'s calculations
varies strongly with frequency with the estimate in equation \ref{eq:EdotWD} corresponding to the upper envelope (see \citealt{Fuller2012} Fig. 8 and eq. 78 or \citealt{Burkart2013} Fig C1). 

Equation \ref{eq:EdotWD} predicts super-Eddington tidal heating rates as $a \rightarrow r_{\rm MT}$ (\citealt{Burkart2013}'s results yield tidal heating rates of similar magnitude).    However the tidal heating is only active for a time $\tgw$ at a given orbital separation so that the total tidal heating at a given $a$ is
\begin{equation}
E_{\rm heat} \approx \dot{E}_{\rm heat} \, \tgw \sim
10^{47} \, {\rm erg} \ f^{-1/5} \, \left(\frac{\MBH}{10^5 M_\odot}\right)^{2/15} \pfrac{a}{r_{\rm MT}}^{-7/5} \pfrac{\amt}{2}^{-7/5} \,.
\label{eq:EheatWD}
\end{equation}
The total dissipated tidal energy prior to $a \sim r_{\rm MT}$ is a small fraction $\sim 0.1 \%$ of the WD binding energy so that only a small fraction of the WD mass can be directly unbound by tidal heating.  

For the large tidal heating rates present when $a \sim r_{\rm MT}$, the internal gravity waves excited by the tidal forcing dissipate by wave breaking near the surface of the star (i.e., when the wave displacement in the radial direction is sufficiently large to overturn the stable stratification).  Inspection of typical WD models shows that for $a \sim r_{\rm MT}$, the waves break in the very outer layers of the star where the exterior mass is $\lesssim 10^{-4} M_\odot$.  Because there is no significant change to the stellar structure in the interior where the tidal internal gravity waves are generated, equations \ref{eq:EdotWD} and \ref{eq:EheatWD} should remain valid even in the presence of tidal heating.    As a result, the most likely outcome of tidal heating in the WD case is a modest outflow generating accretion onto the central BH prior to the onset of unstable mass-transfer.   An upper limit on the mass-loss rate generated is $\dot 2 \dot E_{\rm heat}/v_{\rm esc}^2 \lesssim 10^{-3} M_\odot \, {\rm yr^{-1}}$ (given eq. \ref{eq:EdotWD}) but as argued after equation \ref{eq:EheatWD} the cumulative mass lost is small given the short GW inspiral times.   It is also possible that tidal heating generates runaway fusion of the surface H layer where the tidel heating is concentrated \citep{Fuller2012b}, but this is only the exterior $\sim 10^{-4} M_\odot$ of the WD.   Overall, we conclude that the small mass loss due to tidal heating on circular orbits is unlikely to significantly change the observational signatures estimated in \S \ref{sec:EM_from_wind}.

\subsubsection{Solar-type Stars}
For stars with a finite size radiative core and outer convection zone, internal gravity waves are excited at the radiative-convective boundary and can dissipate by wave breaking in the core of the star where the amplitude of the wave grows due to geometrical convergence \citep{Goodman1998}. Standard calculations (e.g., \citealt{Zahn1977, Goodman1998,Kushnir2017}) of the wave energy flux for a sun-like star on a circular orbit with a small asynchronous rotation due to GW inspiral predict tidal heating rates of
\begin{equation}
\dot E_{\rm heat} \sim
4 \times 10^{30} \, \ergss \left(\frac{\MBH}{10^6 M_\odot}\right)^{11/12 } \pfrac{a}{r_{\rm MT}}^{-85/16} \pfrac{\amt}{2}^{-85/16}
\label{eq:Edotsun}
\end{equation}
where $r_{\rm MT} = \amt \rtidal$ is now for a sun-like star. 
This tidal heating rate is negligible compared to the stellar luminosity and thus tides (for circular orbits) have no significant effect on the stellar structure prior to the onset of unstable mass transfer.   Physically, the reason that the tidal heating rate for the MS star is much less than for a WD is the smaller tidal energy for the MS star (due to the lower stellar binding energy) and the much longer GW inspiral time for a main sequence star approaching tidal disruption.     

Equation \ref{eq:Edotsun} is not applicable to low mass fully convective stars because such stars do not support internal gravity waves.  
However, the resulting tidal heating rate for circular orbits is still negligible due to the general considerations elucidated above (in particular, long GW inspiral times leading to low heating rates for circular inspiral driven by gravitational waves).  We consider fully convective stars on eccentric orbits in the next section.

\subsection{Eccentric Orbits}

For a star on an eccentric orbit with $e \ll 1$ and synchronous spin, the tidal heating rate is approximately 1/3 of the orbital energy transferred from tides to the primary star, with the remaining energy goes into spinning up the star to maintain synchronous rotation \citep[e.g.,][]{Zahn1977}.  We now quantify the magnitude of this tidal heating rate for both white dwarfs and main sequence stars, respectively.   A useful governing equation capturing all of the regimes considered below is that
\begin{equation}
    \dot E_{\rm heat} \sim A \pfrac{E_*}{\tau_{\rm dyn}} \, e^2 \pfrac{a}{r_{\rm MT}}^{-\gamma}  \amt^{-\gamma}
\label{eq:tidegen} 
\end{equation}
where $E_* = GM_*^2/R_*$ is of order the stellar binding energy, $\tau_{\rm dyn}$ is the star's dynamical time, and the dimensionless constants $A \lesssim 1$ and $\gamma \gg 1$ vary somewhat depending on the stellar type and tidal physics.

\subsubsection{White Dwarfs}

For a synchronously rotating white dwarf on a mildly eccentric orbit the tidal heating rate due to the excitation of internal gravity waves is 
\begin{equation}
\dot E_{\rm heat} \sim  3 \times 10^{45} \, {\rm erg \, s^{-1}} \ e^2 \ f \left(\frac{a}{r_{\rm MT}}\right)^{-15} \,
\pfrac{\amt}{2}^{-15} \,.
\label{eq:EdotWDecc}
\end{equation}
In equation \ref{eq:EdotWDecc} we have again used \citet{Fuller2012}'s analytic approximation to the numerical calculation of the tidal excitation rate (their eq. 78) and $f$ is the same dimensionless number discussed in \S \ref{sec:WDe0}.

A corollary of the very large tidal energy input in equation \ref{eq:EdotWDecc} is that the orbit can in principle circularize due to tides at sufficiently small orbital separation.    We find, however, the tidal circularization time is at best comparable to the gravitational wave inspiral time even when $a \simeq r_{\rm MT}$ and so circularization is only a modest effect.

The total heating of the WD during GW inspiral is 
\begin{equation}
E_{\rm heat} \simeq
2 \times 10^{52} \, {\rm erg} \, e^2 \, f\, \left(\frac{\MBH}{10^5 M_\odot}\right)^{-2/3} \pfrac{a}{r_{\rm MT}}^{-11} \pfrac{\amt}{2}^{-11} \,.
\label{eq:EheatWDe}
\end{equation}
If the orbital eccentricity is sufficiently small as $a \rightarrow r_{\rm MT}$ (due to GW decay of the eccentricity) the total tidal heating during GW inspiral is much less than the binding energy of the WD; the critical eccentricity is perhaps $e \sim 0.01$.  In this case it is plausible that the WD remains globally unscathed and mass transfer is primarily initiated when $r \simeq r_{\rm MT}$ (similar to our argument in \S \ref{sec:WDe0}).  For more eccentric orbits the WDs stellar structure will be strongly modified by the cumulative tidal heating.   This could initiate mass transfer at somewhat larger orbital separation by inflating the stellar radius.   Alternatively, it is possible that tidal heating would lead to a thermonuclear runaway and the explosion of the WD.   

\subsubsection{Solar-Type Stars}

For a solar type star with tidal heating set by the excitation of internal gravity waves, the tidal heating rate from standard calculations \citep{Goodman1998} is independent of the companion BH mass\footnote{There is a typo in \citet{Goodman1998} eq. 15; the term $(M_1+M_2)/M_1$ should be raised to the $-5/3$ power.} and is
\begin{equation}
\dot E_{\rm heat}
\simeq 4 \times 10^{38} \, \, {\rm erg \, s^{-1}}
e^2  \pfrac{a}{r_{\rm MT}}^{-23/2} \pfrac{\amt}{2}^{-23/2} \,.
\label{eq:Edotsolarecc}
\end{equation}
Note that this is much smaller than the corresponding heating rate for a WD in the previous section because the tidal heating rate $\propto (G M_*^2/R_*)/\tau_{\rm dyn} \propto R_*^{-5/2}$.   This is a factor of $10^5$ for a WD vs. a main sequence star.\footnote{The actual heating rate in eq. \ref{eq:Edotsolarecc} is yet smaller because it is suppressed by dimensionless factors related to the density at the solar radiative-convective boundary and the size of the solar convection zone.}

For tidal heating to be negligible until mass transfer commences requires $\dot E_{\rm heat} \lesssim L_\odot$, i.e., $e \lesssim 0.003$.   For larger eccentricities, the structure of the star will change in response to the tidal heating.
The total heating of the star during the GW inspiral time is approximately
\begin{equation}
    \dot{E}_{\rm heat} \tgw \approx 10^{52} \, {\rm erg } \; e^2 \, \pfrac{a}{r_{\rm MT}}^{-15/2} \, 
    \pfrac{\amt}{2}^{-15/2} \pfrac{M_\odot}{10^6 M_\odot}^{-2/3}, \label{eq:EMSe}
\end{equation}
valid for $e \ll 1$ (such that $\tgw$ was taken for a circular orbit). We thus conclude that the cummulative effect of tidal heating when $a \gtrsim r_{\rm MT}$ is expected to drastically change the stellar structure when $e \gtrsim 0.01$, with the deposited energy exceeding the star's binding energy.    We also note that the tidal heating rate of Eq.~\ref{eq:Edotsolarecc} implies a circularization timescale at most of order the GW inspiral time and so is reasonable to neglect. 

\subsubsection{Fully Convective Stars}

The estimated tidal heating rate in equation \ref{eq:Edotsolarecc} is not appropriate for a fully convective low mass star which does not support internal gravity waves.  Instead, in such stars, excitation of inertial waves is plausibly the dominant source of tidal dissipation.  \citet{Ogilvie2013} provides an elegant estimate of the tidal frequency-averaged excitation rate of inertial waves by considering the response of a star to an impulsive (in time) tidal force.    \citet{Barker2020} applies this estimate to main sequence low mass stars and shows that the dimensionless tidal quality factor $Q' \simeq 200 (\omega_*/\omega_{\rm rot})^2 \equiv 200 Q'_{200} (\omega_*/\omega_{\rm rot})^2$
for fully convective stars, where $\omega_*$ is the stellar dynamical frequency and $\omega_{\rm rot}$ is its rotation frequency\footnote{$Q'$ is related to the more often-used tidal quality factor $Q$ by $Q=2k_2 Q'/3$ where $k_2$ is the apsidal motion constant.}.  Here we have defined $Q'_{200}$ to encapsulate uncertainty in the tidal excitation rate and variation with stellar mass.  Even in sun-like stars, however, which have comparatively, modest convection zones, $Q'$ is only a factor of $\sim 5$ larger in \citet{Barker2020}'s calculation (see his Fig 4).   Assuming synchronous rotation, this tidal quality factor corresponds to a stellar heating rate of
\begin{equation}
\dot E_{\rm heat}
\sim 3 \times 10^{40} \ \, {\rm erg \, s^{-1}} \,
\pfrac{e^2}{Q'_{200}} \, \pfrac{a}{r_{\rm MT}}^{-21/2} \pfrac{\amt}{2}^{-21/2} \,.
\label{eq:EdotIWecc}
\end{equation}
We stress again that equation \ref{eq:EdotIWecc} averages over the resonant response of  individual inertial waves and is thus only a rough estimate of the heating rate due to inertial waves.   Nonetheless this estimate indicates that the inertial wave heating rate is perhaps $\sim 100$ times larger than that associated with internal gravity waves in solar type stars (eq. \ref{eq:Edotsolarecc}).  This demonstrates that quite circular orbits $e \lesssim 10^{-3}$ are probably necessary for tidal heating to not significantly change the stellar structure prior to the onset of mass-transfer. Most low mass stellar EMRIs will thus have their structure strongly modified by tidal heating at semi-major axes of a few $\rtidal$,  initiating mass transfer somewhat earlier than expected in the absence of tidal heating.

\bibliography{main}{}
\bibliographystyle{aasjournal}

\end{document}